\documentclass[12pt]{article}

\usepackage{amsmath,amsfonts,latexsym}%
\usepackage{fullpage,color}%
\usepackage{graphicx}%
\usepackage{url,hyperref}%
\usepackage{wrapfig}%
\usepackage{theorem}%
\usepackage{xspace}%
\usepackage[active]{srcltx} 
\usepackage{picins}%
\usepackage{paralist}%
\usepackage{ifpdf}%
\usepackage{flowfram}%
\usepackage{boxedminipage}%
\usepackage{wide}%
\usepackage{euscript}

\usepackage{proofapnd}

\renewcommand{\Re}{{\rm I\!\hspace{-0.025em} R}}
\renewcommand{\th}{th\xspace}

\definecolor{blue25}{rgb}{0,0,0.95} \newcommand{\emphic}[2]{%
   \textcolor{blue25}{%
      \textbf{\emph{#1}}}%
   \index{#2}}

\newcommand{\emphi}[1]{\emphic{#1}{#1}}

\newcommand{\rp}{\psi}%
\newcommand{\DAG}{\Term{DAG}\xspace}

\newcommand{\KDS}{\Term{K{D}S}\xspace}
\newcommand{\CHX}[1]{{\mathcal{CH}}\pth{#1}}

\newcommand{\BenThanks}[1]{\thanks{Department of Computer
      Science; 
      University of Illinois; 
      201 N. Goodwin Avenue;
      Urbana, IL, 61801, USA;
      {\tt \si{raichel}2\atgen{}uiuc.edu}; {\tt
         \si{\url{http://www.uiuc.edu/\string~raichel2/}}.} #1}}

\newcommand{\AX}[1]{#1_1}
\newcommand{\BX}[1]{#1_2}

\providecommand{\ds}{\displaystyle}

\newcommand{\complex}{\mathcal{C}}
\newcommand{\complexA}{\AX{\complex}}
\newcommand{\complexB}{\BX{\complex}}

\newcommand{\cell}{\Delta}
\newcommand{\cellA}{\Psi}
\newcommand{\cellB}{\Upsilon}

\newcommand{\myqedsymbol}{\rule{2mm}{2mm}}

\newenvironment{proof}{\trivlist\item[]\emph{Proof}:}%
                  {\unskip\nobreak\hskip 1em plus 1fil\nobreak%
                           \myqedsymbol
                           \parfillskip=0pt%
                           \endtrivlist}

\newcommand{\BallC}{\mathbf{b}}
\newcommand{\BallX}[1]{\mathbf{b}\pth{#1}}

\newtheorem{theorem}{Theorem}[section]%
{\theorembodyfont{\rm} \newtheorem{algorithm}[theorem]{Algorithm}}%
{\theorembodyfont{\rm} \newtheorem{remark}[theorem]{Remark}}%

\newtheorem{observation}[theorem]{Observation}%
\newtheorem{lemma}[theorem]{Lemma}%
\newtheorem{corollary}[theorem]{Corollary}%

{\theorembodyfont{\rm} \newtheorem{defn}[theorem]{Definition}}

\newcommand{\cardin}[1]{\left| {#1} \right|}%
\newcommand{\CH}{{\mathcal{CH}}}%

\newcommand{\sRadius}{\mu}

\newcommand{\simpX}[1]{\mathrm{s{i}m{p}l}\pth{#1}}
\newcommand{\SimpComplexityC}{\mathsf{N}}

\newcommand{\SimpComplexityK}[3]{\SimpComplexityC\pth{#1, #2, \dots,#3}}
\newcommand{\NleqC}{\EuScript{N}}
\newcommand{\Reachable}{\EuScript{R}}
\newcommand{\Nleq}[1]{\NleqC_{\leq #1}}

\newcommand{\relevant}{reachable\xspace}

\newcommand{\resemblanceX}[1]{\ensuremath{#1}-relative free space complexity\xspace}
\newcommand{\deciderFr}{\Algorithm{decider}\xspace}
\newcommand{\approxBinarySearch}{\Algorithm%
   {search{}Events}\xspace}
\newcommand{\approxDistances}{\Algorithm{approx{}Distances}\xspace}
\newcommand{\Pairwise}[1]{\binom{#1}{2}}

\newcommand{\intervalFr}{\Algorithm{search{}Interval}\xspace}
\newcommand{\approxMean}{\Algorithm{a{}p{}r{}x{}Mean{}}\xspace}
\newcommand{\VertexSet}[1]{{V}\pth{#1}}
\newcommand{\sEvents}{Z}
\newcommand{\solver}{\Algorithm{solver}\xspace}

\newcommand{\curveA}{\pi}

\newcommand{\curveAs}{\pi'}
\newcommand{\curveB}{\sigma}

\newcommand{\curveC}{\tau}

\newcommand{\R}{\mathbb{R}}

\newcommand{\distFr}[2]{\mathsf{d}_{\EuScript{F}}\pth{#1, #2}}
\newcommand{\distFrM}[2]{\mathsf{d}_{\EuScript{F}}^\mathrm{m} \pth{#1, #2}}
\newcommand{\distFrW}[2]{\mathsf{d}_{\EuScript{F}}^\mathrm{w} \pth{#1, #2}}
\providecommand{\pth}[2][\!]{#1\left({#2}\right)}
\newcommand{\widthX}[2]{{\mathrm{width}}_{#2}\pth{#1}}
\newcommand{\distCmd}[1]{\left\| {#1}  \right\|}
\newcommand{\distX}[2]{\distCmd{#1 - #2}}
\newcommand{\Frechet}{Fr\'{e}chet\xspace}
\newcommand{\Family}{\EuScript{F}}
\providecommand{\MakeBig}{\rule[-.2cm]{0cm}{0.4cm}}
\providecommand{\MakeSBig}{\rule[0.0cm]{0.0cm}{0.35cm}} 
\newcommand{\distMean}[2]{\mathsf{d}_{mean} \pth{#1, \dots,#2}}

\newcommand{\PntSet}{\mathsf{P}}
\newcommand{\relX}[2][\!]{\mathrm{rel}\pth[#1]{#2}}

\newcommand{\elevX}[1]{\mathrm{elev}\pth{#1}}
\newcommand{\DistSets}[2]{\mathsf{d}\pth{#1, #2}}

\newcommand{\Vertices}[1]{V\pth{#1}}
\newcommand{\Edges}[1]{E\pth{#1}}
\newcommand{\etal}{\textit{et~al.}\xspace}

\newcommand{\CellPathExt}[2]{\mathrm{cell_{#1}}\pth{#2}}
\newcommand{\CellPathA}[1]{\mathrm{cell_{\curveA}}\pth{#1}}
\newcommand{\CellPathB}[1]{\mathrm{cell_{\curveB}}\pth{#1}}
\newcommand{\CellPath}[2]{\mathrm{cell}_{#1,#2}}

\newcommand{\signX}[1]{\mathrm{sign}\pth{#1}}
\newcommand{\Term}[1]{\textsf{#1}}
\newcommand{\MST}{\Term{MST}\xspace}

\newcommand{\itemref}[2][]{(#1\ref{item:#2})}
\newcommand{\thmlab}[1]{{\label{theo:#1}}}
\newcommand{\thmref}[1]{Theorem~\ref{theo:#1}}

\DefineNamedColor{named}{RedViolet} {cmyk}{0.07,0.90,0,0.34}

\newcommand{\AlgorithmI}[1]{{\textcolor[named]{RedViolet}{\texttt{\bf{#1}}}}}
\newcommand{\Algorithm}[1]{{\AlgorithmI{#1}\index{algorithm!#1@{\AlgorithmI{#1}}}}}

\newcommand{\sv}{s}
\newcommand{\tv}{t}

\newcommand{\sA}{\sv_1}
\newcommand{\sB}{\sv_2}
\newcommand{\tA}{\tv_1}
\newcommand{\tB}{\tv_2}

\newcommand{\dcA}{\complex_1}
\newcommand{\dcB}{\complex_2}

\newcommand{\decider}{\Algorithm{decider}\xspace}
\newcommand{\extractI}{\Algorithm{extract}\xspace}
\newcommand{\compFr}{\Algorithm{comp{}Fr}\xspace}
\newcommand{\Interval}{\mathcal{I}}

\newcommand{\FDleq}[1]{F_{\leq{#1}}}
\newcommand{\minRad}[1]{r_{\mathrm{min}}\pth{#1}}

\newcommand{\sPnt}{s}%
\newcommand{\ePnt}{t}%
\newcommand{\perimX}[1]{\mathrm{perim}\pth{#1}}

\newcommand{\edge}{\mathsf{e}}

\newcommand{\pnt}{\mathsf{p}}
\newcommand{\pntA}{\mathsf{q}}

\newcommand{\pntC}{\mathsf{v}}
\newcommand{\Graph}{\mathsf{G}}
\providecommand{\si}[1]{#1}
\newcommand{\eps}{{\varepsilon}}%

\newcommand{\seclab}[1]{\label{sec:#1}}
\newcommand{\secref}[1]{Section~\ref{sec:#1}}

\newcommand{\obslab}[1]{\label{observation:#1}}
\newcommand{\obsref}[1]{Observation~\ref{observation:#1}}

\newcommand{\remlab}[1]{\label{remark:#1}}
\newcommand{\remref}[1]{Remark~\ref{remark:#1}}

\newcommand{\alglab}[1]{\label{alg:#1}}
\newcommand{\algref}[1]{Algorithm~\ref{alg:#1}}

\newcommand{\figlab}[1]{\label{fig:#1}}
\newcommand{\figref}[1]{Figure~\ref{fig:#1}}

\providecommand{\deflab}[1]{\label{def:#1}}

\newcommand{\lemlab}[1]{\label{lemma:#1}}
\newcommand{\lemref}[1]{Lemma~\ref{lemma:#1}}

\newcommand{\itemlab}[1]{\label{item:#1}}%
\newcommand{\sep}[1]{\,\left|\, {#1} \MakeBig\right.}
\newcommand{\brc}[1]{\left\{ {#1} \right\}}
\newcommand{\pbrc}[2][\!\!]{#1\left[ {#2} \MakeBig \right]}

\newcommand{\SarielThanks}[1]{\thanks{Department of Computer Science;
      University of Illinois; 201 N. Goodwin Avenue; Urbana, IL,
      61801, USA; {\tt sariel\atgen{}uiuc.edu}; {\tt
         \url{http://www.uiuc.edu/\string~sariel/}.} #1}}
\newcommand{\atgen}{\symbol{'100}}

\title{The \Frechet Distance Revisited and Extended%
   \thanks{The latest full version of this paper is available online
      \cite{hr-fdre-11-manuscript}.}}

\author{%
   Sariel Har-Peled%
   \SarielThanks{Work on this paper was partially supported by a NSF
      AF award CCF-0915984.}%
   \and%
   Benjamin Raichel%
   \BenThanks{Work on this paper was partially supported by a NSF
      AF award CCF-0915984.}%
}

\date{\today}

\begin{document}

\maketitle

\begin{abstract}
    Given two simplicial complexes in $\Re^d$, and start and end vertices in each
    complex, we show how to compute curves (in each complex) between
    these vertices, such that the \Frechet distance between these
    curves is minimized.  As a polygonal curve is a complex, this
    generalizes the regular notion of weak \Frechet distance between
    curves.  We also generalize the algorithm to handle an input of
    $k$ simplicial complexes.

    Using this new algorithm we can solve a slew of new problems, from
    computing a mean curve for a given collection of curves, to
    various motion planning problems.  Additionally, we show that for
    the mean curve problem, when the $k$ input curves are $c$-packed, 
    one can $(1+\eps)$-approximate the mean curve in near linear time,
    for fixed $k$ and $\eps$.

    Additionally, we present an algorithm for computing the 
    strong \Frechet distance between two curves, which is simpler than 
    previous algorithms, and avoids using parametric search.
\end{abstract}

\section{Introduction}

The \Frechet distance provides a way to measure the similarity between
curves.  Unlike the Hausdorff distance, which treats the curves as
sets, the \Frechet distance takes into account the structure of the
curves, by requiring continuous reparameterizations of the curves.
Informally, the \Frechet distance between two curves, $\curveA$ and
$\curveB$, is the minimum length leash needed to walk a dog when
the person walks along $\curveA$ and the dog walks along $\curveB$. 

In this paper, we are interested in extending this concept to
facilitate solving more general motion planning problems.

\paragraph{Previous Work.}

The \Frechet{} distance and its variants have been used to measure
similarity between curves in applications such as dynamic time-warping
\cite{kp-sudtw-99}, speech recognition \cite{khmtc-pgbhp-98},
signature and handwriting recognition \cite{mp-cdtw-99, skb-fdba-07},
matching of time series in databases \cite{kks-osmut-05}, as well as
geographic applications, such as map-matching of vehicle tracking data
\cite{bpsw-mmvtd-05,wsp-anmms-06}, and moving objects analysis
\cite{bbg-dsfm-08, bbgll-dcpcs-08}.

Alt and Godau \cite{ag-cfdbt-95} showed how to compute the \Frechet
distance between two polygonal curves in $\Re^d$, of total complexity
$n$, in $O(n^2\log n)$ time\footnote{In their paper, as well as ours, 
$d$ is considered to be a constant.}.  It is an open problem to find a
subquadratic algorithm for computing the \Frechet distance for two
curves. The decision problem (i.e., deciding whether the \Frechet
distance is smaller than a given value) has a lower bound of $\Omega(n
\log n)$ \cite{bbkrw-wtd-07}.  Driemel \etal \cite{dhw-afdrc-10}
provided a $(1+\eps)$-approximation for polygonal curves, that works
in $O(N(\eps,\curveA,\curveB) + N(1,\curveA,\curveB) \log n)$ time,
where $N(\eps,\curveA,\curveB)$ is the relative free space complexity of two
curves under simplification.  In particular, their algorithm runs in
$O\pth{ cn/\eps + cn\log n }$ time for $c$-packed curves.

One can generalize the problem to consider an input of multiple
curves.  Dumitrescu and Rote \cite{dr-fdsc-04} consider the problem of
simultaneously minimizing the \Frechet distance between all pairs of a
set of $k$ curves.  They show one can get a 2-approximation in
$O(n^2\log n)$ time, whereas the naive extension to $k$ curves of the
exact algorithm takes $O(n^k\log n)$ time.  Buchin et
al. \cite{bbklsww-mt-10} consider the problem of finding the median
trajectory (polygonal curve) of a set of $k$ trajectories in the plane
that all share the same starting and ending points, where the median
trajectory is a trajectory contained in the union of the $k$ input
trajectories that must cross at least half of the input trajectories
in order to reach the unbounded face (though \Frechet distance does
not directly come in to play in their problem, it is related to the
mean curve problem that we consider in \secref{mean}).

The notion of the \Frechet distance can also be generalized to
encompass distances between surfaces.  Unfortunately, for general
surfaces the decision problem is NP-hard \cite{g-cms-99}.  In fact,
whether the \Frechet distance for general surfaces is computable is
still an open problem.  Recently Alt and Buchin \cite{ab-csbs-10}
showed that the problem is semi-computable between surfaces, and
polynomial time computable for the weak \Frechet distance. The problem
is hard even if the surfaces are well-behaved terrains, see Buchin
\etal \cite{bbs-fdsss-10}.

Moving away from \Frechet distances between surfaces, Alt \etal
\cite{aerw-mpm-03} presented an $O(n^2 \log^2 n)$ time algorithm to
compute the \Frechet distance between two graphs.  Specifically, they
require that one of the two graphs has to be entirely traversed and in
the other graph we seek the path that minimizes the \Frechet distance
to the path of this traversal.

\paragraph{Complexes.}
The notion of a \emphi{complex} (which is an abstract simplicial
complex together with its realization), defined formally in
\secref{complex}, is a generalization of polygonal curves,
triangulations, meshes, straight line graphs, etc.  In particular, our
algorithm uses complexes as inputs and as such would apply for all
these different inputs in a verbatim fashion.

\paragraph{Our Contribution.}

Given two complexes and start and end points in each one of them, we
present a general algorithm that computes the two curves in these
complexes that are closest to each other, under the weak \Frechet distance,
and connect the corresponding start and end points. In fact, the resulting 
curves that our algorithm produces are locally optimal (see \remref{local}).  
The running time of this new algorithm is $O\pth{ n^2 }$ (where $n$ is the 
complexity of the input complexes). Our algorithm
can be interpreted as an extension of the algorithm of Alt and Godau
\cite{ag-cfdbt-95} for computing the weak \Frechet distance between
polygonal curves. \cite{ag-cfdbt-95} and subsequent work on the \Frechet 
distance has relied heavily on using the parametric space of the input 
complexes.  Our main contribution is the usage of the product 
complex instead of the parametric space -- this seemingly small 
change in viewpoint enables us to easily encode the, potentially 
very complicated, connectivity information of the input complexes 
in a simple way.  This in turn allows us to observe that the classical 
\Frechet distance techniques in fact apply to a wide array problems (many 
of which had not been previously considered) involving higher 
complexity input complexes.  The notion of product complexes is implicit 
in previous work on this topic \cite{aerw-mpm-03, bpsw-mmvtd-05,wsp-anmms-06}.  
However, by bringing this concept to the forefront, we get a clean framework 
that facilitates our new applications.

As concrete applications of our algorithm consider the following
variants, all of them immediately solvable by our algorithm:
\medskip
\begin{compactenum}[(A)]
    \item \textbf{\Frechet for paths with thickness.} Imagine the
    classical setting of the \Frechet distance where a person walks a
    dog, but both the dog and the person might walk on
    paths that have non-zero width. That is, the input is two simple
    polygons (i.e., ``thickened'' paths) and one needs to compute the
    two paths of minimum \Frechet distance between them that lie
    inside their respective polygons.

    \item In a similar vane, consider a wiring problem: You are given
    a three dimensional model (of say a car or an airplane) specified by
    its mesh, and you are given a rough suggested path connecting two
    points in the mesh. Our algorithm can compute the optimal wiring
    path inside the model that is closest, under the \Frechet
    distance, to the suggested rough path.
\end{compactenum}

\medskip

Interestingly, this approach also extends to inputs of more than two
complexes, and also to arbitrary convex functions between
these different complexes. Specifically, consider a situation where
the input includes $k$ complexes $\complex_1, \ldots, \complex_k$. The
reader might think about the complex $\complex_i$ as the domain of the
$i$\th agent.  Given a location in each of these complexes of their
respective agent (i.e., a point $\pnt_i$ inside the complex
$\complex_i$ and the simplex $\cell_i\subseteq \complex_i$ that
contains it) consider a scoring function $f(\pnt_1, \ldots, \pnt_k)$
that assigns a cost to the configuration $(\pnt_1,\ldots,
\pnt_k)$. Furthermore, assume that this scoring function is convex on
the domain $\cell_1 \times \cell_2 \times \cdots \times \cell_k$, and
this holds for any combination of such simplices.  Now, given that
the agents want to move from some starting vertices $v_1, \ldots, v_k$
to ending vertices $v_1',\ldots, v_k'$, the new algorithm can
compute the synced motion of these $k$ agents from the starting
configuration to the ending configuration, such that the maximum cost
of any configuration used throughout the motion is minimized.

The reader might consider these settings a bit abstract, so here are a
few examples of problems that can be solved using this framework:
\medskip
\begin{compactenum}[(P1)]    
    \item \itemlab{i:1} Mean curve. Given a set of $k$
    curves in $\Re^d$, find a new curve that minimizes the maximum 
    \Frechet distance between this new curve and each of the input
    curves. Namely, this computes a mean curve for a given
    collection of curves.

    \item One can compute the optimal way to walk $k$ agents on $k$
    curves/complexes such that the maximum distance between any pair of agents,
    at any point in time, is minimized.

    \item Compute the optimal way for the $k$ agents to walk on the
    $k$ curves/complexes, such that the maximum average distance
    between any pair of agents is minimized (the average is over all
    pairs).

    \item \itemlab{i:penultimate} Walk a pack of dogs while minimizing 
    a weighted sum of the leash lengths (i.e. maybe some dogs
    need to be kept close since they like to chase squirrels).

    \item \text{Motion minimizing the perimeter of the convex hull.}
    Given $k$ curves/complexes that $k$ agents have to move on (in the
    plane), compute a motion from the start points to the end points,
    such that the maximum perimeter of the convex hull is minimized
    throughout the motion.
\end{compactenum}
\bigskip

The running time of all these algorithms for $k$ input
complexes of total complexity $n$ is $O\pth{ n^k }$.  In 
\secref{pack} we show that by making minor (realistic) assumptions 
about the input curves, for the median curve problem, one can remove 
the exponential dependence on $k$ (however, the constant retains the 
exponential dependence on $k$).  

\medskip

As a side problem, we also consider the problem when the input is two
\DAG complexes, which are directed acyclic straight line graphs
embedded in $\Re^d$.  By considering the product space of two such
complexes (instead of the parametric space) we show that the decision
problem can be solved in $O\pth{ n^2 }$ time.  We then present a simple
randomized technique to solve the general problem in $O(n^2 \log n)$ time. In
particular, this provides an alternative algorithm that computes the (strong)
\Frechet distance between two polygonal curves without using
parametric search. Specifically, this algorithm is considerably
simpler than the algorithm of Alt and Godau \cite{ag-cfdbt-95}, while
matching its running time. Previous efforts to avoid the parametric
search by using randomization resulted in algorithms that are slower
by a logarithmic factor \cite{ov-psmp-04, cw-gfdis-09}. This new
algorithm uses ideas applied for the problem of slope selection
\cite{m-roass-91} to the computation of the \Frechet distance.  See
\thmref{f:r:DAG} for details.

\paragraph{Organization.}
In \secref{preliminaries}, we define the \Frechet distance and
complexes formally, as well as introduce the key concept of using 
the product space instead of the parametric space, when defining
the free space.  \secref{alg} outlines the main algorithm of the paper, where
it is shown that by applying the convexity property of the free
space, our problem can be converted into the problem of computing the 
bottleneck shortest path.  We also generalize the algorithm to handle
$k$ input complexes, as well as arbitrary convex functions.  In
\secref{applications} we outline some applications of the main
algorithm.  In \secref{pack} we show that when the $k$ input curves
are $c$-packed, one can solve the mean curve problem in near linear time.  
In \secref{DAG:complexes}, we present an algorithm for	
computing the monotone \Frechet distance between two  curves or
between two \DAG complexes.

\section{Preliminaries}
\seclab{preliminaries}

\subsection{Curves and the \Frechet Distance}

Let $\curveA \subseteq \Re^d$ be a curve; that is, a continuous mapping 
from $[0,1]$ to $\Re^d$. In the following, we will identify $\curveA$ 
with its range $\curveA\pth{[0,1]} \subseteq \Re^d$ if it is clear from 
the context.

A \emphi{reparameterization} is a continuous one-to-one function
$f:[0,1]\rightarrow [0,1]$, such that $f(0)=0$ and $f(1)=1$. Given two
reparameterizations $f$ and $g$ for two curves $\curveA$ and
$\curveB$, respectively, define their \emphi{width} as
\[
\widthX{\curveA,\curveB}{f,g} = \max_{s \in [0,1]}
\distX{\curveA(f(s))}{\curveB(g(s))}.
\]
This can be interpreted as the maximum length of a leash one needs to
walk a dog, where the dog walks along $\curveA$ according to $f$,
while the handler walks along $\curveB$ according to $g$. In this
analogy, the \Frechet distance is the shortest possible leash
admitting such a walk.  Formally, given two curves $\curveA$ and
$\curveB$ in $\Re^d$, the \emphi{monotone \Frechet distance} between them is 
\[
\distFr{\curveA}{\curveB} = \min_{%
   \substack{f:[0,1] \rightarrow [0,1]\\
      g:[0,1]\rightarrow [0,1]}} \widthX{\curveA,\curveB}{f,g},
\]
where $f$ and $g$ are orientation-preserving reparameterizations of
the curves $\curveA$ and $\curveB$, respectively. We will also be 
interested in the \emphi{weak \Frechet distance}, where the
reparameterizations are required to be continuous but not necessarily
bijections (i.e., one is allowed to walk backwards on their respective curve).
In our problem we will be defining the curves in the respective domains.  Hence 
finding curves that minimize the weak \Frechet distance and finding curves 
that minimize the strong \Frechet distance, are equivalent problems.  The 
reader should note, however, that when the input domains are curves, our algorithm 
is equivalent to computing the weak \Frechet distance between those curves.

\subsection{Complexes}
\seclab{complex}

An $n$-dimensional \emphi{simplex} is the convex hull of n+1 affinely
independent vertices.  We call the convex hull of any m+1 vertex
subset of the vertices of a simplex, an $m$-dimensional
\emphi{subcell} (or face) of that simplex (note that a subcell is in
fact an $m$-dimensional simplex). A proper subcell is one such that
$m<n$.

An \emphi{abstract simplicial complex} $\complexA=(P,\Family)$, is a
set system.  The elements of $P$ are \emphi{points} and the elements
of $\Family$ are subsets of $P$ called \emphi{simplices}.  An abstract
simplicial complex is downward closed; that is for any $\cellA\in
\Family$, and $\cellB\subseteq \cellA$, it holds that $\cellB\in
\Family$.  For our purposes, the ground set $P$ will always be a
subset of $\Re^d$.  We also use the natural \emphi{realization} of the 
abstract simplicial complex $(P,\Family)$, by mapping any simplex
$\cellA\in \Family$ to $\relX{\cellA} = \CHX{ \cellA }$, where
$\CHX{\cellA }$ denotes the convex hull of $\cellA$.  Throughout our
discussion we assume that for any $\cellA\in \Family$, we have
$|\cellA| = \dim(\CHX{\cellA}) + 1$ (i.e. $\cellA$ is affinely
independent).  We also require that our realization is locally
consistent; that is $\forall \cellA,\cellB \in \Family$, if
$\cellA\cap \cellB\neq \emptyset$ then $\relX{\cellA} \cap
\relX{\cellB} = \relX{\cellA \cap \cellB }$.

Note, that the geometric realization of such an abstract simplicial
complex does not induce a simplicial complex.  For example, such an
abstract simplicial complex might define a self intersecting polygonal
curve, where two disjoint simplices $\cellA$ and $\cellB$ have that
$\relX{\cellA }$ and $\relX{\cellB}$ intersect in their interior.  In
the following, we will refer to an abstract simplicial complex
together with its realization as a \emphi{complex}.

For a complex, $\complexA$, we will refer to any simplex in
$\complexA$ as a \emphi{cell} of $\complexA$.  The dimension of a
complex is the maximum dimension of any of its cells.  We say
$\cellA\in \complexA$ is a maximal cell of $\complexA$ if there is no
$\cellB\in \complexA$ such that $\cellA\subset \cellB$ (note that a
maximum cell is one such that $dim(\cellA)=dim(\complexA)$).

A pair of simplices $\cellA,\cellB$ are \emphi{adjacent} if
$\cellA\subseteq \cellB$ or $\cellB\subseteq \cellA$.  A
\emphi{simplicial path} in a complex is a function $\phi: [0,1]
\rightarrow \Family$, such that:
\begin{inparaenum}[(A)]
    \item For any $\cell \in \Family$, we have that
    $\phi^{-1}\pth{\cell}$ is a finite union of open intervals and
    points.
    \item If $\phi(\cdot)$ has only two distinct values (say $\cell$ and
    $\cellA$) on an interval $[x,y] \subseteq [0,1]$, then the
    simplices $\cell$ and $\cellA$ are adjacent.
\end{inparaenum}

A curve $\curveA \subseteq \Re^d$ parameterized over $[0,1]$ is a 
realization of a simplicial path $\phi$, if for any $t\in [0,1]$ we have
that $\curveA(t) \in \relX{\phi(t)}$ and $\phi(t)$ is the simplex of lowest
dimension of $\Family$ that contains $\curveA(t)$.  
In our applications, a maximal interval $(x,y)$
such that $\phi$ is constant corresponds to a
straight segment of $\curveA$.  In particular, when dealing with a curve
$\curveA \subseteq \Re^d$, we will assume that its associated 
simplicial path is also known.

In the following we will abuse notation and refer to $\cellA$ as a
shorthand for $\relX{\cellA}$.  In particular, for a point $p\in
\Re^d$, we will say that $p$ is in the simplex $\cellA$ if $p\in 
\relX{\cellA}$.

\subsection{Product Spaces}
Let $\complexA=(\PntSet_1,\Family_1)$ and
$\complexB=(\PntSet_2,\Family_2)$ be two simplicial complexes in
$\Re^d$.  Consider the product space $\complexA \times \complexB$.
Intuitively, we view the product space as a subset of the space
$\Re^{2d}$, where the first $d$ coordinates are from $\complexA$ and
the remaining $d$ coordinates are from $\complexB$.  With this view,
$\complexA\times \complexB$ is similar to a simplicial complex
although the cells will be convex polyhedra instead of just
simplices (in the literature this is known as a \emphi{polyhedral complex}).
We define a cell $(\cellA,\cellB)$ of $\complexA\times \complexB$ to
be the product of any cell $\cellA$ from $\complexA$ with any cell
$\cellB$ from $\complexB$.  Its realization is the set
$\relX{\cellA,\cellB } = \relX{\cellA} \times \relX{\cellB}$.  In the
polyhedral complex $\complexA\times \complexB$, two cells $(\cellA,\cellB)$
and $(\cellA ',\cellB ')$ are \emphi{adjacent} if $\cellA$ is adjacent
to $\cellA '$ in $\complexA$ and $\cellB = \cellB '$, or $\cellA =
\cellA'$ and $\cellB$ is adjacent to $\cellB'$ in $\complexB$.  Also,
note that $\complexA\times \complexB$ is connected since, by
assumption, the complexes $\complexA$ and $\complexB$ are connected.

Let $\curveA$ and $\curveB$ be curves with reparameterizations $f$ and
$g$, respectively. Let $\CellPathA{\cdot}$ and $\CellPathB{\cdot}$ be
the simplicial paths associated with $\curveA(f)$ and $\curveB(g)$, respectively.  Since
the Cartesian product of two continuous functions is continuous, we
have that $h(t) = (\curveA(f(t)),\curveB(g(t)))$ defines a curve
$\curveC = \cup_t h(t)$ in $\complexA\times \complexB$, which we call
the \emphi{product curve} of $\curveA(f)$ and $\curveB(g)$.  The curve
$\curveC$ has a corresponding \emphi{product cell path} which is the
function $\CellPath{\curveA}{\curveB}\pth{t} = \pth{ \CellPathA{t},
   \CellPathB{t}}$.  (For the sake of simplicity of exposition, we are
assuming here that $\CellPathA{t}$ and $\CellPathB{t}$ do not change
their value simultaneously at the same time $t$.)

For two complexes $\complexA$ and $\complexB$ in $\R^d$, and a
parameter $\delta \geq 0$, consider a cell $(\AX{\cell},\BX{\cell})$
in $\complexA\times \complexB$.  For a point $\pnt = \pth{\AX{\pnt},
   \BX{\pnt}} \in \pth{\AX{\cell},\BX{\cell}}$, its \emphi{elevation}
is the quantity $\elevX{\pnt} = \elevX{ \AX{\pnt}, \BX{\pnt} } =
\distX{\AX{\pnt}}{\BX{\pnt}}$.  The feasible region in the cell 
$\AX{\cell} \times \BX{\cell}$ is the set
\[
\FDleq{\delta}(\AX{\cell},\BX{\cell}) = \brc{ (x,y)\in \R^{2d} \sep{
      \begin{array}{l}
          x\in \relX{\AX{\cell}} \subseteq \R^d\\
          y\in \relX{\BX{\cell}}\subseteq \R^d\\
          \elevX{{\AX{\pnt}, \BX{\pnt}}} \leq \delta
      \end{array}
   }}.
\]
The feasible region for $\complexA \times \complexB$ (which we will
refer to as the \emphi{free space}\footnote{Note that here the free space 
is defined in terms of the product space, instead of the usual parametric 
space.})  is the set $\FDleq{\delta}(\complexA,\complexB) =
\cup_{\AX{\cell} \in \complexA, \BX{\cell} \in \complexB}
\FDleq{\delta}(\AX{\cell},\BX{\cell})$.

\begin{observation}
    Let $\curveA$ and $\curveB$ be paths in $\complexA$ and
    $\complexB$, respectively, and let $f$ and $g$ be
    reparameterizations of $\curveA$ and $\curveB$ respectively, that
    realize the value $\delta$ of the \Frechet distance.  The product
    curve, $\curveC$, is contained in $F_{\leq \delta}
    (\complexA,\complexB)$. Indeed, for any $t\in [0,1]$, we have
    $\elevX{{\curveA(f(t)),\curveB(g(t))}} \leq \delta$, since $f$ and
    $g$ realize the \Frechet distance between $\curveA$ and $\curveB$.

    \obslab{free}
\end{observation}

\begin{observation}
    Consider a curve $\curveB$ in $\complexA \times \complexB$, such
    that for any point $\pnt \in \curveB$ we have that $\elevX{\pnt}
    \leq \delta$. Then, the projection of this curve into the
    corresponding curves in $\complexA$ and $\complexB$ results in two
    curves $\AX{\curveB}$ and $\BX{\curveB}$ such that
    $\distFr{\AX{\curveB}}{\BX{\curveB}} \leq \delta$.

    Formally, for $t \in[0,1]$, let $\curveB(t) = \pth{
       \AX{\curveB}(t), \BX{\curveB}(t)} \in \complexA \times
    \complexB$ be a parameterization of $\curveB$, and let
    $\CellPathExt{\curveB}{t} = \pth{\CellPathExt{\AX{\curveB}}{t},
       \CellPathExt{\BX{\curveB}}{t} }$ be its associated product cell
    path, such that for any $t$ we have $\curveB(t) \in \relX{
       \CellPathExt{\curveB}{t}}$. Clearly, $\AX{\curveB}(t)$ and
    $\BX{\curveB}(t)$ are parameterized curves in the complexes
    $\complexA$ and $\complexB$, respectively. Furthermore, for any
    $t\in [0,1]$, we have that
    $\distX{\AX{\curveB}(t)}{\BX{\curveB}(t)} = \elevX{\curveB(t)}
    \leq \delta$. As such, $\distFr{\AX{\curveB}}{\BX{\curveB}} \leq
    \delta$.

    \obslab{encode}
\end{observation}

\subsection{Convexity of the Free Space in a Cell}

We need the following straightforward result.  
We include the proof for the sake of completeness.

\newcommand{\LemmaBodyConvex}{%
   Let $F_{\leq \delta} = F_{\leq \delta}(\complexA,\complexB)$ be the
   free space of the complexes $\complexA$ and $\complexB$, both
   contained in $\Re^d$, for some $\delta\geq 0$.  Then $F_{\leq \delta}(\cellA,\cellB) =
   F_{\leq \delta} \cap \pth{ \relX{\cellA}\times \relX{\cellB}}$ is a
   convex set, for any cell $(\cellA, \cellB)$ of $\complexA\times
   \complexB$.
   
   Putting it differently, the elevation function $\elevX{\cdot }$ is
   convex over $\relX{\cellA} \times \relX{\cellB}$, for any cell
   $(\cellA, \cellB)$ of $\complexA\times \complexB$.  %
}

\begin{lemma}
    \LemmaBodyConvex

    \lemlab{convex}
\end{lemma}

\begin{proof}
    Let $\cellA$ and $\cellB$ be simplices in $\complexA$ and
    $\complexB$, respectively, and let $F=F_{\leq
       \delta}(\cellA,\cellB)$.  By the definition of free space, we 
    know that $F$ is just the sublevel set (i.e. the level set and
    everything less than that level) of the function $h:\Re^d\times
    \Re^d \rightarrow \Re$, where $h(u,v) = \distX{u}{v}$, when
    applied to $\cellA\times \cellB$.  It is known that the sublevel
    set of a convex function with a convex domain, is convex.  Hence
    all we need to show is that $h$ is convex (note that the domain is
    convex since $\cellA$ and $\cellB$ are convex).
    
    So let $u,u' \in \cellA$ and $v, v'\in \cellB$.  We show that $t
    h(u,v)+(1-t)h(u'v') \geq h\pth{\MakeSBig t(u,v) +(1-t)(u',v')}$,
    for $t\in [0,1]$.  Equivalently, we show that the function
    $\widehat{h}(t) = h\pth{\MakeSBig t(u,v) +(1-t)(u',v')}$ is convex
    on the interval $[0,1]$, i.e. $\widehat{h}(t)\leq
    (1-t)\widehat{h}(0)+(t)\widehat{h}(1)$ (actually we need to prove
    such an inequality for all choices of $u,u'\in \cellA$ and
    $v,v'\in \cellB$, which holds since they were
    chosen arbitrarily).

    Expanding out this function we get,
    \begin{align*}
        \widehat{h}(t)%
        &=%
        h\pth{\MakeSBig t(u,v) +(1-t)(u',v') }%
        =%
        h\pth{\MakeSBig u'+t(u-u'),v'+t(v-v')}\\
        &%
        =%
        \distCmd{u'+t(u-u') - v'- t(v-v')}%
        =%
        \distCmd{(u'-v')+t(u+v'-u'-v)}.
    \end{align*}
    
    Hence $\widehat{h}(t)$ is just the equation for the distance
    between a point on a linearly parameterized line and the origin.
    We have by \lemref{line:point} below that this function is convex and so
    we are done.
\end{proof}

\begin{lemma}
    The function representing the distance between a point on a
    linearly parameterized line $l(t)$ and the origin, is a convex
    function.  Specifically, let $a$ and $b$ be vectors in
    $\Re^d$, then the function $f(t) = \distCmd{a+tb}$, is
    convex.

    \lemlab{line:point}
\end{lemma}

\begin{proof}
    We know that $f(t)$ is of the form,
    \begin{align*}
        f(t) &= \sqrt{\sum_i (a_i+t b_i)^2} = \sqrt{ \alpha t^2 +
           \beta t + \gamma},
    \end{align*}
    where $\alpha$, $\beta$, and $\gamma$ are some constants such that
    $\alpha t^2 +\beta t + \gamma$ is non-negative.  By the helper
    lemma below, however, we know such a function is convex.
\end{proof}

\begin{lemma}
    Consider the quadratic function $\alpha t^2 + \beta t + \gamma$,
    where $\alpha, \beta$ and $\gamma$ are some constants such that
    the function is non-negative. Then, the function $f(t) = \sqrt{
       \alpha t^2 + \beta t + \gamma}$ is convex.
\end{lemma}
\begin{proof} 
    Since $\alpha t^2 + \beta t + \gamma \geq 0$ for all $t$, it must
    be that $\alpha > 0$, and the corresponding quadratic formula
    either has no roots, or a single root, which implies that $\beta^2
    -4\alpha \gamma \leq 0$. Now,
    \[
    f'(t) = \frac{2\alpha t + \beta}{2 \sqrt{ \alpha t^2 + \beta t +
          \gamma}} = \frac{h(t)}{f(t)},
    \]
    for $h(t) = \alpha t + \beta/2$. Similarly,
    \[
    f''(t) %
    =%
    \frac{f(t)h'(t) - f'(t) h(t)}{(f(t))^2} =%
    \frac{\alpha f(t) - (h(t))^2/f(t)}{(f(t))^2}  =%
    \frac{(f(t))^2 - (h(t))^2/\alpha}{(f(t))^3/\alpha}.
    \]
    
    Now, since $f(t)$ is always non-negative, we have that
    \begin{align*}
        \signX{f''(t)} &= \signX{(f(t))^2 - (h(t))^2/\alpha} = \signX{
           \alpha t^2 + \beta t + \gamma - \alpha t^2 - \beta t -
           \beta^2 /4\alpha} %
        \\& = %
        \signX{\gamma -\beta^2 /4\alpha } = %
        \signX{4\alpha\gamma -\beta^2 } \geq 0,
    \end{align*}
    since $\alpha > 0$ and $\beta^2 -4\alpha \gamma \leq 0$.
\end{proof}

\subsection{Bottleneck Shortest Path Algorithm}
As a subroutine to our main algorithm, we need the following algorithm
to compute bottleneck shortest paths in linear time, which is accepted as 
folklore, but we include for sake of completeness.

Let $\Graph=(V,E)$ be an undirected graph, with weight function $w$ on the edges.  For a given path $p$ in $\Graph$, let $b(p)=\max_{e\in E(p)} w(e)$. 
We call a path connecting two vertices $s$ and $t$ in $\Graph$ an \emphi{$st$ path}.   We say that an $st$ path, $p'$, is a \emphi{bottleneck shortest path}, if among the set of all $st$ paths, $P$, we have that $b(p')=\min_{p\in P} b(p)$.

\begin{lemma}
\lemlab{bottleneck}
Let $\Graph=(V,E)$ be an undirected graph, with weight function $w$ on the edges.  For any pair of vertices $s,t \in V$, one can compute in $O(|V|+|E|)$ time the bottleneck shortest path between $s$ and $t$.
\end{lemma} 

\begin{proof}
In the following, let $p$ denote the bottleneck shortest path between $s$ and $t$.
First we check to see if $s$ and $t$ are connected in $\Graph$ and output $\infty$ if not.  Otherwise, the algorithm proceeds in recursive stages.  In each stage, we compute the median weight edge, $e_{med}$, and consider the graph $\Graph_{\leq med}=(V,E_{\leq med})$, where $E_{\leq med} \subseteq E$ consists of all the edges of weight less than or equal to the median.  We then compute the connected components of $\Graph_{\leq med}$.  If $s$ and $t$ are in the same component of $\Graph_{\leq med}$ then $b(p)\leq w(e_{med})$.  In this case we recurse on the graph $\Graph_{\leq med}$ since none of the larger weight edges can be in $p$.  Otherwise, $b(p)> w(e_{med})$ and so we can contract each connected component of $\Graph_{\leq med}$ down to a vertex, and recurse on the graph with these vertices and the edges in $E_{>med} =E\setminus E_{\leq med}$.  In either case, at the end of a stage we remove all isolated vertices (this is also done before the first stage).  Eventually, we will reach a stage with some small constant amount of edges in which case we can then solve the problem by brute force.

Let $m_i$  and $n_i$ be the number of edges and vertices, respectively, in the $i$th stage.  
Computing the connected components, finding the median, and removing bad edges and isolated vertices all take $O(n_i+m_i)$ time (or better).  Since we always remove isolated vertices at the end of a stage, we know that $n_i=O(m_i)$, and hence each stage takes $O(m_i+n_i)=O(m_i)$ time.   In each stage we delete (roughly) half the edges and so there are $O(\lg m)$ stages, and in each stage $O(m_i) = O(2^{-i}m)$ work is done, and so O(m+n) work is done in total (the $n$ is included since the first check for isolated vertices will cost $O(n)$ time, and it might be that $m=o(n)$). 
\end{proof}


\section{Computing Optimal \Frechet Paths in Complexes}
\seclab{alg}

We are given as input two complexes, $\complexA$ and $\complexB$,
along with corresponding start and end vertices $s_1$, $t_1$ and
$s_2$, $t_2$.  We wish to compute the paths $\curveA$ and $\curveB$ in
$\complexA$ and $\complexB$, respectively, that minimize the \Frechet
distance over all paths that start and end at the respective start and
end vertices.  

\subsection{Algorithm}
We construct a graph $\Graph = (V,E)$, called the \emphi{cell graph}
of $\complexA\times \complexB$.  Specifically, each cell
$(\AX{\cell},\BX{\cell})$ of $\complexA\times \complexB$ corresponds
to a vertex $v_{(\AX{\cell},\BX{\cell})} \in V$, and for every pair
$v_{(\AX{\cell}, \BX{\cell})}, v_{(\AX{\cell}', \BX{\cell}')} \in V$
we create an edge iff $(\AX{\cell}, \BX{\cell})$ and $(\AX{\cell}',
\BX{\cell}')$ are adjacent in $\complexA \times \complexB$.  For
$\AX{\cell} \in \complexA$ and $\BX{\cell} \in \complexB$, the
\emphi{elevation} of their corresponding vertex $v = v_{(\AX{\cell},
   \BX{\cell})} \in V$ is $\elevX{v} =
\DistSets{\AX{\cell}}{\BX{\cell}}$, where
$\DistSets{\AX{\cell}}{\BX{\cell}} = \min_{\pnt\in
   \relX[]{\AX{\cell}}, \pntA \in \relX[]{\BX{\cell}}} \elevX{{\pnt,
      \pntA}}  = \min_{\pnt\in \relX[]{\AX{\cell}}, \pntA \in
   \relX[]{\BX{\cell}}} \distX{\pnt}{\pntA}$ is the distance between
these simplices. The point realizing this minimum is the
\emphi{realization} of the vertex $v$, and is denoted by $\relX{v}$.

The cell graph is clearly connected since
$\complexA\times \complexB$ is connected. As such, for any pair of
vertices $u,v \in \Vertices{\Graph}$ there exists a $uv$ path in
$\Graph$. The \emphi{elevation} of a path $\rho$, denoted by
$\elevX{\rho}$, is the maximum elevation of any vertex in $\rho$.  The
\emphi{lowest} $uv$ path in $\Graph$ is the $uv$ path with minimum elevation.

We compute the lowest $st$ path in $\Graph$ (where $s=(s_1, s_2)$ and
$t=(t_1, t_2)$), in order to determine the desired curves with minimum \Frechet
distance.  To this end, we set the elevation of any edge $uv \in
E(\Graph)$ to be $\elevX{uv} =\max\pth{\elevX{u}, \elevX{v}}$.  Clearly computing 
the lowest $st$ path in a weighted graph is the same as computing the bottleneck 
shortest path, and so using the algorithm of \lemref{bottleneck} we can efficiently compute the lowest $st$ path, 
$\rho =v^1 \ldots v^m$, in $\Graph$.  We return the polygonal path $\relX{v^1} \relX{v^2} \cdots \relX{v^m}
\subseteq \complexA\times \complexB$ as the desired curve (which by
\obsref{encode} encodes the two desired curves and their
reparameterizations).

\subsection{Analysis}

\subsubsection{Correctness}

As the following lemmas show, the cell graph captures the relevant
information for our problem.

\begin{lemma}
    Let $\complexA$ and $\complexB$ be two complexes, and let $s_1$
    and $t_1$ be vertices of $\complexA$ and let $s_2$ and $t_2$ be
    vertices of $\complexB$.  Then, if there exists an $s_1t_1$ path
    $\curveA$, in $\complexA$, and an $s_2t_2$ path $\curveB$, in
    $\complexB$, such that $\distFr{\curveA}{\curveB} = \delta$ then
    there exists a $v_{(s_1,s_2)}v_{(t_1,t_2)}$ path, $\rho$, in
    $\Graph(\complexA,\complexB)$ such that $\elevX{\rho} \leq \delta$.

    \lemlab{to:graph}
\end{lemma}

\begin{proof:in:appendix}{\lemref{to:graph}}
    Let $f$ and $g$ be the reparameterizations of $\curveA$ and
    $\curveB$, respectively, that achieve the value $\delta$ for the
    \Frechet distance.  By \obsref{free} the product curve $\curveC =
    \bigcup_t \pth{\MakeSBig \curveA(f(t)),\curveB(g(t)) }$, defines a
    path in $\complexA \times \complexB$ from $(s_1,s_2)$ to
    $(t_1,t_2)$ that is contained in the free space $F_{\leq
       \delta}(\complexA,\complexB)$.  Let
    $\CellPath{\curveA}{\curveB}\pth{t}$ be the product cell path in
    $\complexA \times \complexB$ that corresponds to $\curveC(t)$.
    Naturally, the value of $\CellPath{\curveA}{\curveB}\pth{t}$
    corresponds to a vertex in $\Graph$, and let $v(t)$ denote this
    vertex.  It is easy to verify that the sequence of different
    vertices visited by $v(t)$, as $t$ increases from $0$ to $1$, is a
    valid path in $\Graph$. Indeed, a product cell path defines a
    sequence of adjacent cells of $\complexA\times \complexB$ as $t$
    increases from $0$ to $1$, which corresponds to a path 
    $\rho = v^1, \ldots, v^m$ in $\Graph$.

    Observe, that for any $t\in [0,1]$, we have that 
    \begin{align*}
        \elevX{v(t)} = \elevX{
           v_{\CellPath{\curveA}{\curveB}\pth{t}}}%
        =%
        \min_{\substack{\pnt \in {\CellPathA{t}}, \\\pntA \in {\CellPathB{t}}}}
        \distX{\pnt}{\pntA}%
        \leq%
        \distX{\curveA(f(t))}{\curveB(g(t))} \leq \delta.
    \end{align*}
    As such, $\elevX{\rho} = \max_i \elevX{v^i} =
    \max_{t} \elevX{v(t)} \leq \delta$.
\end{proof:in:appendix}

\begin{lemma}
    Let $\complexA$ and $\complexB$ be two complexes, and let $s_1$
    and $t_1$ be vertices of $\complexA$ and let $s_2$ and $t_2$ be
    vertices of $\complexB$.  Then, if there exists a
    $v_{(s_1,s_2)}v_{(t_1,t_2)}$ path $\rho$ in $\Graph(\complexA,\complexB)$ such that
    $\elevX{\rho} = \delta$ then there exists an $s_1t_1$ path,
    $\curveA$, in $\complexA$ and an $s_2t_2$ path, $\curveB$, in
    $\complexB$, such that $\distFr{\curveA}{\curveB}= \delta$.

    \lemlab{back:from:graph}
\end{lemma}

\begin{proof:in:appendix}{\lemref{back:from:graph}}
    Let $\rho=v^1\ldots v^m$, where $v^1= v_{(s_1,s_2)}$ and $v^m=
    v_{(t_1,t_2)}$.  Each vertex $v^i$ in $\rho$ corresponds to a pair
    of cells $\cell^i = (\AX{\cell^i},\BX{\cell^i})$, where
    $\AX{\cell^i} \in \complexA$ and $\BX{\cell^i} \in \complexB$.
    Furthermore, for every $i$, there exists two points $\AX{\pnt^i}
    \in \AX{\cell^i}$ and $\BX{\pnt^i}\in \BX{\cell^i}$, such that
    $\elevX{\pnt^i} = \distX{\AX{\pnt^i}}{\BX{\pnt^i}} =
    \DistSets{\AX{\cell^i}}{\BX{\cell^i}}$, where $\pnt^i =
    \pth{\AX{\pnt^i}, \BX{\pnt^i}}$. 

    Observe, that for all the vertices of the polygonal path $Z =
    \pnt^1 \pnt^2 \ldots \pnt^m$, we have that $\elevX{\pnt^i} =
    \DistSets{\AX{\cell^i}}{\BX{\cell^i}} = \elevX{v^i} \leq
    \elevX{\rho} = \delta$. As such, all the vertices of $Z$ are in
    the free space $\FDleq{\delta}$.

    For any $i$, the $i$\th segment of $Z$ is $\pnt^i \pnt^{i+1}$. It
    corresponds to the edge $v^i v^{i+1}$ in the graph $\Graph$, which
    connects adjacent cells in $\complexA \times \complexB$. In
    particular, it must be that either $\cell^i \subseteq \cell^{i+1}$
    or $\cell^{i+1} \subseteq \cell^{i}$. Assume the latter happens
    (the other case is handled in a symmetric fashion). We have that 
    $\pnt^i\pnt^{i+1} \subseteq \cell^{i}$. Furthermore, by the
    convexity of the free space inside a single cell (i.e.,
    \lemref{convex}), we have that $\pnt^i\pnt^{i+1} \subseteq \cell^i
    \cap \FDleq{\delta}$. We conclude that $Z \subseteq \FDleq{\delta}$.
    Since the two endpoints of $Z$ are 
    $\pth{ s_1, s_2} = \pnt^1$ and 
    $\pth{ t_1, t_2} = \pnt^m$, $Z$ corresponds to the desired paths 
    $\curveA$ and $\curveB$ such that $\distFr{\curveA}{\curveB}= \delta$. 
\end{proof:in:appendix}


\begin{corollary}
    Let $\complexA$ and $\complexB$ be two complexes, and
    let $s_1$ and $t_1$ be vertices of $\complexA$ and let $s_2$ and $t_2$
    be vertices of $\complexB$.  Moreover, let $\curveA$ and $\curveB$
    be the paths in $\complexA$ and $\complexB$, respectively, that
    minimize the \Frechet distance over all pairs of $s_1t_1$ and
    $s_2t_2$ paths.  Then we have that $\distFr{\curveA}{\curveB}=
    \delta$ if and only if the lowest $v_{(s_1,s_2)}v_{(t_1,t_2)}$
    path, $\rho$, in $\Graph(\complexA,\complexB)$ has $\elevX{\rho}=
    \delta$.
\end{corollary}

\subsubsection{Running Time Analysis}

Computing the lowest $st$ path takes $O(|V|+|E|)$ time by \lemref{bottleneck}.  
Since a vertex in the cell graph represents a pair
of simplices from $\complexA$ and $\complexB$, we know that
$\cardin{\Vertices{\Graph}}=O(|\complexA||\complexB|)$.  We also know
that $\cardin{\Edges{\Graph}} = O\pth{ \cardin{\Vertices{\Graph}}}$
since each cell in $\complexA\times \complexB$ has at most $O(1)$
proper subcells (specifically $O\pth{ 2^{2d}} = O(1)$).  Hence the
running time of the algorithm is $O\pth{ n^2 }$, where
$n = \max\pth{ \cardin{\complexA}, \cardin{\complexB}}$.

Putting everything together, we get the following result.

\begin{theorem}
    Let $\complexA$ and $\complexB$ be two simplicial complexes, and
    $n=\max(|\complexA|,|\complexB|)$.  Given any pair of start and
    end vertices from $\complexA$ and any pair of start and end
    vertices from $\complexB$, we can compute, in $O\pth{ n^2
    }$ time, the paths $\curveA$ and $\curveB$ in $\complexA$ and
    $\complexB$, respectively, that minimize the \Frechet distance
    over all paths that start and end at the respective start and end
    vertices.

    \thmlab{2:complexes}
\end{theorem}

\begin{remark}
    It is easy to verify that \thmref{2:complexes} yields a path that
    is locally as low as possible. Formally, if the solution in the
    polyhedral complex is a curve $\curveA$, then for any subcurve
    $\curveB\subseteq \curveA$, we have the property that for any
    other curve $\curveC$, that has the same endpoints of $\curveB$,
    it holds that $\elevX{\curveC} \geq \elevX{\curveB}$.

    When computing the \Frechet distance for two curves for example,
    this property implies that the parameterization we get is never
    lazy -- it always tries to be as tight as possible at any given
    point in time.
\remlab{local}
\end{remark}

\subsubsection{Applications}

\paragraph{\Frechet for paths with thickness.}
Given two polygons (maybe with holes) in the plane and start and
end vertices in the two polygons, one can triangulate the two polygons
and then feed them into \thmref{2:complexes}. This results in two
paths in the two triangulations that minimize the \Frechet distance
between the paths. As a concrete application, this can be used for
solving the classical \Frechet distance problem where the input curves
have thickness associated with them and one can move in this enlarged
region. Indeed, each ``thickened'' curve can be represented as a
polygon, and hence we can apply the above algorithm.

\paragraph{Wiring.} 
The wiring problem, mentioned in the introduction, can be solved by
immediate plug and play into the above result.

\paragraph{Motion planning in planar environments.}
Consider the case where you need to plan the motion of two entities in a
two dimensional environment, where they have to stay close together
(i.e., \Frechet distance) while complying with different constraints
on which part of the environment they can travel on. As a concrete
example, one entity might be a pedestrian and the other might be a
vehicle. The pedestrian can not use the road, and the vehicle can not
use the sidewalk or the parks available. Finding the best motion for
the two entities is no more than solving the \Frechet problem in this
setting. Indeed, we compute a triangulation of the environment for the 
first entity, and then remove all triangles and edges that can not be 
used by the first entity. Similarly, we compute a triangulation for the 
second entity, removing the regions that are unusable for it.

Now, applying the algorithm of \thmref{2:complexes} to these two
triangulations (with the desired starting and ending points) results
in the desired motion.

\medskip
Naturally, the algorithm of \thmref{2:complexes} can be applied in
more general settings where the input is three dimensional, etc.


\subsection{Generalized Algorithm for $k$ Complexes}


Let us recap the algorithm from the previous section. We considered
finding the path in the product space (of two complexes) such that the
maximum value of $f(x,y) = \distX{x}{y}$ among all the points $(x,y)$
in the path is minimized.  If we add an extra dimension for the value
of $f$, then one can think of $f$ as defining a terrain. Then the
problem becomes computing a path that does not traverse high in this
terrain.  The free space was the sublevel set of $f$ for some
parameter $\delta$.  Next, we defined the elevation of a vertex in the
cell graph to be the minimum value of $f$ for the cell that the vertex
corresponds to.  By observing that $f$ was a convex function within
each cell in the product space, we were able to argue that the value
of the best path (i.e. lowest maximum value of $f$) was equivalent to
the value of the bottleneck shortest path, and thus the problem was 
efficiently solvable.

With this abstract description, the only property of $f$ that we used
was that it was convex within each cell in the product space.  Hence,
we can conclude that the same procedure will work for any choice of
$f$, so long as it is convex within each cell in the product space.

We can generalize the problem even further.  Earlier we considered
only two complexes.  However, there is no reason why we can not
consider an input of $k$ complexes, for some arbitrary integer $k$.
In order to handle this case we generalize all our earlier definitions
for two complexes in the following natural way.

Let $\complex_1 = (\PntSet_1, \Family_1), \ldots$, $\complex_k =
(\PntSet_k, \Family_k)$ be a set of $k$ simplicial complexes in
$\R^d$.  Consider the product space $\complex_1 \times \cdots \times
\complex_k$.  Intuitively, we view the product space as a subset of
the space $\R^{kd}$.  We define a cell $(\cell_1,\ldots, \cell_k)$ of
$\complex = \complex_1\times \cdots \times \complex_k$ to be the product of $k$
cells, where $\cell_i \in \complex_i$, for $i=1, \ldots, k$.  Its
realization is the set $\relX{{\cell_1,\ldots, \cell_k}} =
\relX{\cell_1} \times \ldots\times \relX{\cell_k}$.  In
$\complex_1\times \ldots \times \complex_k$, two cells $(\cell_1,
\ldots, \cell_k)$ and $(\cellA_1, \ldots, \cellA_k)$ are
\emphi{adjacent} if there is a $j$ such that for all $i\neq j$,
$\cell_i=\cellA_i$ and $\cell_j$ is adjacent to $\cellA_j$ in
$\complex_j$.

We now are given a function $f$ defined over $\Re^{kd}$ that is convex
for any cell $\relX{\cell_1,\ldots, \cell_k}$. As before, we build the
cell graph $\Graph$ of the polyhedral complex $\complex$. Every vertex $v$ of $\Graph$
corresponds to a cell $\cell$ of $\complex$, and its elevation is the
minimum value of $f$ in this cell.

As before, we are given start vertices $s_1, \ldots, s_k$ and end
vertices $t_1,\ldots, t_k$ in these $k$ complexes. We compute the
lowest elevation path between the vertex in $\Graph$ corresponding to
$(s_1, \ldots, s_k)$ and the vertex in $\Graph$ corresponding to
$(t_1,\ldots, t_k)$.  Arguing as before, it is easy to show that the
resulting path in the graph can be realized by a path in $\complex$
that yields the $k$ desired paths and their reparameterizations.  As
such, we get the following result.

\begin{theorem}
    We are given $k$ simplicial complexes $\complex_1, \ldots,
    \complex_k$, $n=\max_i \cardin{\complex_i}$, start vertices $s_1
    \in \complex_1, \ldots s_k \in \complex_k$, end vertices $t_1 \in
    \complex_1, \ldots, t_k \in \complex_k$, and a function $f:
    \relX{\complex} \rightarrow \Re$ that is convex for any cell in
    the realization of $\complex = \complex_1 \times \cdots \times
    \complex_k$.

    Then, one can compute, in $O\pth{n^k }$ time, $k$ curves
    $\curveA_1, \ldots, \curveA_k$ (and their reparameterizations
    $\rp_1, \ldots, \rp_k$) connecting $s_1, \ldots, s_k$ to $t_1,
    \ldots, t_k$, respectively, such that $\max_t f(\MakeSBig
       \curveA_1(\rp_1(t)),\allowbreak \ldots,\allowbreak \curveA_k(\rp_k(t)) )$ is
    minimized, among all such curves and reparameterizations.

    \thmlab{k:complex}
\end{theorem}

\section{Applications}
\seclab{applications}


\subsection{Mean Curve}
\seclab{mean}

We are given $k$ polygonal curves $\curveA_1,\ldots, \curveA_k$ in
$\Re^d$, and we would like to compute a curve $\curveB$ that minimizes the
maximum weak \Frechet distance between $\curveB$ and each one of the curves
$\curveA_1,\ldots, \curveA_k$.

For a set of points $\PntSet \subseteq \Re^d$, let 
$\minRad{\PntSet}$ denote the radius of the minimum enclosing ball of
$\PntSet$. 

\begin{lemma}
    Let $\PntSet(t)$ be a set of points in $\Re^d$ moving linearly
    with $t$. Then, the function $\minRad{t} =\minRad{\PntSet(t)}$ is
    convex.
    \lemlab{obvious}
\end{lemma}

\begin{proof:in:appendix}{\lemref{obvious}}
    Fix any three times, $x<y<z$, where $y=\alpha x+(1-\alpha)z$ for
    some $\alpha\in (0,1)$.  Let $\pnt_i(t)$ denote the $i$\th moving
    point of $\PntSet(t)$.

    Let $\pntC_x$ (resp. $\pntC_z$) be the center of the
    minimum enclosing ball of $\PntSet(x)$ (resp. $\PntSet(z)$), and let
    $\pntC(y) = \alpha \pntC_x  + (1-\alpha)\pntC_z$.
    Observe that
    \begin{align*} 
        \minRad{y} &= \minRad{\MakeSBig \PntSet(\alpha x +
           (1-\alpha)z) } %
        \leq%
        \max_{i} \distX{\pntC(\alpha x + (1-\alpha)z)}{\pnt_i(\alpha x
           + (1-\alpha)z)} \\&%
        \leq%
        \max_i \pth {\alpha \distX{\pntC_x}{\pnt_i(x)} + (1-\alpha)
           \distX{\pntC_z}{\pnt_i(z)}}%
        \\&%
        =%
        \alpha \minRad{x} + (1-\alpha)\minRad{z},
    \end{align*}
    since the distance between a pair of linearly moving points is
    convex (for example by \\ \lemref{line:point}).
\end{proof:in:appendix} 

\medskip \noindent
Using the lemma above, we get the following desired result.

\begin{lemma}
    Given $k$ curves $\curveA_1, \ldots, \curveA_k$ in $\Re^d$ with
    total complexity $n$, one can compute, in $O \pth{ n^k }$ 
    time, a curve $\curveB$ that minimizes $\max_{i}
    \distFrW{\curveA_i}{\curveB}$, where
    $\distFrW{\curveA_i}{\curveB}$ is the weak \Frechet distance
    between $\curveA_i$ and $\curveB$.

    \lemlab{easy:2}
\end{lemma}

\begin{proof:in:appendix}{\lemref{easy:2}}
    A cell in the polyhedral complex of $\curveA_1 \times \cdots \times
    \curveA_k$ is the product of $k$ segments (or points) in
    $\Re^d$. For a point $\pnt = (\pnt_1, \ldots, \pnt_k) \in
    \Re^{dk}$ inside such a cell, consider the elevation of $\pnt$ to
    be $f(\pnt ) =\minRad{\brc{\pnt_1, \ldots, \pnt_k}}$.
    \lemref{obvious} implies that $f(\cdot)$ is convex inside each
    such cell. As such, applying \thmref{k:complex} to the given
    curves, using the function $f(\cdot)$, results in a
    parameterization that minimizes the maximum radius of the minimum
    enclosing ball throughout the motion. Since the center of the
    minimum enclosing ball (for continuously moving points) changes
    continuously over time, the curve formed by this center throughout
    the motion is a natural mean curve. Let $\curveB$ denote this
    curve. It is easy to prove that the maximum \Frechet distance of
    $\curveB$ to any of the curves $\curveA_1, \ldots, \curveA_k$ is
    the minimum such value among all possible curves.
\end{proof:in:appendix}

\subsection{Walking a Pack of Dogs}

So suppose you have a pitbull, a chiwawa, a corgi, and a terrier.  You
want to walk all the dogs at the same time instead of walking each one
individually.\footnote{Since clearly you are a person that is very
   concerned with efficiency.}  However, as before, long leashes are
expensive, so you want to minimize the maximum length leash (among all
the leashes) that you need to use.

Formally, you are given $k$ complexes, $\complex_1,\ldots,
\complex_k$, and start and end vertices $s_i, t_i \in \complex_i$, for
$i=1,\ldots,k$. The first complex corresponds to the person leading
the dogs, and the complexes $\complex_2,\ldots, \complex_k$
corresponds to the $k-1$ given dogs.  You wish to find the set of
paths, $\curveA_1, \ldots, \curveA_k$, and corresponding
reparameterizations, $\rp_1, \rp_2, \ldots, \rp_k$, such that,
\[
\max_{t \in [0,1]} \max_{i>1}
\distX{\curveA_1(\rp_1(t))}{\curveA_i(\rp_i(t))} = \max_{t\in [0,1]}
f(\curveA_1(\rp_1(t)), \ldots, \curveA_k(\rp_k(t))),
\]
is minimized, where $f\pth{\pnt_1, \ldots, \pnt_k} = \max_{i}
\distX{\pnt_1}{\pnt_i}$.

\begin{lemma}
    Given $k$ polygonal curves $\curveA_1, \ldots, \curveA_k$ of
    total complexity $n$, one can compute non-monotone reparameterizations of these
    curves such that $\max_t \max_i
    \distX{\curveA_1\pth{\rp_1\pth{t}}}{\curveA_i\pth{\rp_i\pth{t}}}$
    is minimized. The running time of the algorithm is
    $O\pth{n^k }$.

    This works verbatim for complexes, and in this case the algorithm
    also computes the paths inside the complexes realizing the
    \Frechet distance.

    \lemlab{walk:many:dogs}
\end{lemma}

\begin{proof:in:appendix}{\lemref{walk:many:dogs}}
    We need to prove that the function $f\pth{\pnt_1, \ldots, \pnt_k} = \max_{i}
    \distX{\pnt_1}{\pnt_i}$ is convex within each cell
    in order to apply \thmref{k:complex}.

    So, consider a cell $\cell = \pth{\cell_1,\ldots, \cell_k} \in
    \complex = \complex_1 \times \cdots \times \complex_k$. Its realized cell
    $\relX{\cell} = \relX{\cell_1} \times \cdots \times
    \relX{\cell_k}$ is a convex set.  In particular, consider the
    functions of the form $f_i(\pnt_1, \pnt_i) =
    \distX{\pnt_1}{\pnt_i}$, defined over $\relX{\cell_1} \times
    \relX{\cell_i}$, for $2\leq i\leq k$. Each of these functions are
    convex by \lemref{convex} on the domain $\relX{\cell_1} \times
    \relX{\cell_i}$. In particular, setting $g_i(\pnt_1, \ldots,
    \pnt_k) = f_i(\pnt_1, \pnt_i)$, for $i=1,\ldots, k$, results in
    $k$ convex functions over $\relX{\cell}$.

    Clearly, $f\pth{\pnt_1, \ldots, \pnt_k} = \max_{i} g_i\pth{\pnt_1,
       \ldots, \pnt_k}$, which is convex as the maximum of a set of
    convex functions is a convex function.  As such, plugging this
    into \thmref{k:complex} implies the result.
\end{proof:in:appendix}

\subsection{More General Settings}

From the previous example, consider the person and the dogs at any
given time as vertices in space.  The leashes are thus edges
connecting the vertices.  Hence in the above example the topology of
the graph is that of star graphs (i.e. the person is at the center and
the dogs are the ends of the star).  The ``weight'' of each edge in
the graph is the value of a convex function between the respective
pair of vertices at a given instance of time (i.e. the distance of the
person to a specific dog at a specific time).  The general function we
were trying to minimize was the maximum value over the functions
between each pair of vertices.  We were able to conclude that the
overall function was convex because the maximum value of a set of
convex functions, is a convex function.

Let the above described graph be called a \emphi{dependency graph}.
In general we can consider any topology for the dependency graph.
More formally, between every pair of complexes we define a convex
function (note that the zero function is convex, and so we can ignore
certain pairs if we like).  For our global function we can then take
any function of these functions, which preserves convexity.  For
example, taking the maximum, the sum, or (positively) weighted sum of
convex functions is again a convex function.  Therefore, all of the
applications \itemref[P]{i:1}--\itemref[P]{i:penultimate} mentioned in
the introduction are solvable immediately within this framework.

\subsubsection{Minimizing Perimeter of Motion}

We are given $k$ complexes $\complex_1, \ldots, \complex_k$ all with
realizations in the plane. As before, we are given $k$ starting
vertices $\sPnt_1, \ldots, \sPnt_k$ and $k$ ending vertices $\ePnt_1,
\ldots, \ePnt_k$, in these $k$ complexes, respectively. We are
interested in computing the $k$ polygonal paths (and their
reparameterizations) connecting these endpoints, such that the maximum
perimeter is minimized. As before, to use the framework, we need to
show that the perimeter function is convex inside a cell of the
resulting polyhedral complex.  So, consider two points $\pnt = (\pnt_1,
\ldots, \pnt_k)$ and $\pntA = (\pntA_1, \ldots, \pntA_k)$. We need to
show that the perimeter function
\[
\perimX{t} = \perimX{  t\pnt + (1-t)\pntA} = \mathrm{perimeter}
\pth{ \MakeBig\! \CH \pth{  \brc{ \MakeSBig t\pnt_1 + (1-t)\pntA_1,
         \ldots, t\pnt_k + (1-t)\pntA_k}}}
\]
is convex. This fact, which we state below as a lemma, is proved in
\cite{ac-atcfmap-10} using the Cauchy-Crofton inequality.

\begin{lemma}[\cite{ac-atcfmap-10}]
    The perimeter of a set of linearly moving points in the plane is
    a convex function.
\end{lemma}

This implies that the perimeter function is convex inside each cell of 
$\complex = \complex_1\times \cdots \times \complex_k$, and hence the
framework applies.  We thus get the following result.

\begin{lemma}
    Given $k$ complexes $\complex_1, \ldots, \complex_k$ all with
    realizations in the plane, and $k$ starting vertices $\sPnt_1, \ldots,
    \sPnt_k$ and ending vertices $\ePnt_1, \ldots, \ePnt_k$, in
    these $k$ complexes, respectively, then one can compute paths in
    these complexes, and their corresponding reparameterizations, such
    that the maximum perimeter of the moving points during this motion
    is minimized over all such motions. The running time of the
    algorithm is $O( n^k )$.
\end{lemma}

The running time stated above is under the assumption
that computing the minimum perimeter for $k$ points whose
locations are restricted by a cell of the polyhedral complex, can be done in
constant time. This constant would depend on $k$, naturally.

\section{Computing the Mean Curve for c-packed Curves}
\seclab{pack}

Driemel et al. \cite{dhw-afdrc-10} introduced a realistic class of curves, called 
$c$-packed curves.  We now show that when the $k$ input curves are 
$c$-packed, one can compute a $(1+\eps)$-approximation to the mean curve in 
$\widetilde{O} (n\log n)$ time, where $\tilde{O}()$ is used to emphasize that the 
constant depends on $\eps$ and $c$, and exponentially on $k$ and $d$ (see 
\lemref{algCpackTheorem} and \lemref{complexity:l:e:q} for details).  This is a significant improvement
over the algorithm for the general case, presented in \secref{mean}, where the
running time is $O(n^k)$.

In this section, when we refer to the free space, it is meant 
with respect to the mean curve distance function.  In particular, for 
$k$ curves $\curveA_1,\dots,\curveA_k$ let $\distMean{\curveA_1}{\curveA_k}$ 
denote the maximum distance of the mean curve to the $\curveA_i$'s, for 
the optimum reparameterizations.

\subsection{Preliminaries}
\subsubsection{Definitions and Lemmas}
We first cover the definitions and lemmas from \cite{dhw-afdrc-10} 
that are relevant to our problem.

\begin{defn}
    For a parameter $c>0$, a curve $\curveA$ in $\Re^d$ 
    is \emphi{$c$-packed} if for any
    point $\pntA$ in $\Re^d$ and any radius $r > 0$, the
    total length of the portions of $\curveA$ inside the ball $\BallX{\pntA,r}$ is at
    most $c r$.
\end{defn}

\begin{algorithm}
    Given a polygonal curve $\curveA= \pntA_1 \pntA_2 \pntA_3 \ldots
    \pntA_k$ and a parameter $\sRadius>0$, consider the following
    simplification algorithm: First mark the initial vertex $\pntA_1$
    and set it as the current vertex. Now scan the polygonal curve
    from the current vertex until it reaches the first vertex
    $\pntA_i$ that is in distance at least $\sRadius$ from the current
    vertex. Mark $\pntA_i$ and set it as the current vertex. Repeat
    this until reaching the final vertex of the curve, and also mark
    this final vertex.  Consider the curve that connects only the
    marked vertices, in their order along $\curveA$. We refer to the
    resulting curve $\curveAs = \simpX{\curveA,\sRadius}$ as being the
    \emphi{$\sRadius$-simplification} of $\curveA$. Note, that this
    simplification can be computed in linear time.

    \alglab{simplification:algorithm}
\end{algorithm}

We need the following useful facts about $\sRadius$-simplifications from \cite{dhw-afdrc-10}.

\begin{lemma} 
    \begin{compactenum}[(i)]
		\item
      For any curve $\curveA$ in $\Re^d$, and $\sRadius > 0$, we have
    	that $\distFr{\MakeSBig\curveA}{\simpX{\curveA,\sRadius}} \leq
    	\sRadius$.
		
		\item 
    	Let $\curveA$ be a $c$-packed curve in $\Re^d$, let $\sRadius > 0$ be a
    	parameter, and let $\curveAs = \simpX{\curveA, \sRadius}$ be the
    	simplified curve. Then, $\curveAs$ is a $6c$-packed curve.
	\end{compactenum}

    \lemlab{simplification:distance}
\end{lemma}

\begin{observation}
    Let $\curveA$ and $\curveB$ be two given curves, and let $\curveA '$ 
    and $\curveB '$ be their $\mu$ simplified curves, for some value 
    $\mu$.  By \lemref{simplification:distance}, $\distFr{\curveA}{\curveA '}\leq \mu$ 
    and $\distFr{\curveB}{\curveB '}\leq \mu$.  Hence we have 
    reparameterizations $f$ and $g$ such that $\distX{\curveA(f(t))}{\curveA'(t)}\leq \mu$ 
    and $\distX{\curveB(g(t))}{\curveB'(t)}\leq \mu$ for all $t\in [0,1]$ 
    (without loss of generality we can assume these reparameterizations 
    are bijective).  Let $\distFrW{\curveA '}{\curveB '}=\delta$, where $\distFrW{\cdot}{\cdot}$ is the weak \Frechet distance.  Then 
    we have that $\distFrW{\curveA}{\curveB} \leq \delta +2\mu$, 
    since we can just map each pair $(x,y)\in (\curveA ', \curveB ')$ 
    that is seen in the optimal (not necessarily injective) reparameterizations 
    of $\curveA '$ and $\curveB '$, to the corresponding pair in 
    $(\curveA, \curveB)$ determined by $f$ and $g$.  In particular, 
    this implies that for curves $\curveA_1,\dots,\curveA_k$ with 
    corresponding $\mu$ simplifications $\curveA'_1,\dots,\curveA'_k$, 
    we have that $\distMean{\curveA_1}{\curveA_k}\leq \distMean{\curveA'_1}{\curveA'_k}+2\mu$.
	\obslab{strong:weak}
\end{observation}

Let $\curveA_1,\dots,\curveA_k$ be $k$ given curves.  The \emphi{complexity} 
of the \relevant free space for these curves, for a distance $\delta$, 
denoted by $\Nleq{\delta}(\curveA_1,\dots,\curveA_k)$, is the total
number of cells in the polyhedral complex with non-empty intersection with 
$F_{\leq \delta}(\curveA_1,\dots,\curveA_k)$ such that there exists a path with
elevation $\leq \delta$ from the start vertex to that cell.  

\begin{defn}
    For $k$ curves $\curveA_1,\dots, \curveA_k$, let
    \[
    \SimpComplexityK{\eps}{\curveA_1}{\curveA_k} = \max_{\delta \geq 0}
    \;\Nleq{\delta} \pth{\MakeSBig \simpX{\curveA_1, \eps \delta}, \dots,
       \simpX{\curveA_k, \eps \delta }}
    \]
    be the maximum complexity of the \relevant free space for the
    simplified curves.  We refer to
    $\SimpComplexityK{\eps}{\curveA_1}{\curveA_k}$ as the
    \emphi{\resemblanceX{\eps}}of $\curveA_1, \dots, \curveA_k$.
   
    \deflab{simplification:complexity}
\end{defn}

\subsubsection{Subroutines}
We now list the relevant subroutines from \cite{dhw-afdrc-10}, which carry over directly for our problem.

Using the same procedure as in \cite{dhw-afdrc-10}, one can build 
a decider, $\decider(\delta, \eps, \curveA_1,\dots, \curveA_k)$ 
that runs in $O\pth{ \SimpComplexityK{\eps}{\curveA_1}{\curveA_k} }$ time (the only difference 
being that in our case the BFS ignores monotonicity).  Specifically, 
we have the following.

\begin{lemma}
    Let $\curveA_1, \dots, \curveA_k$ be $k$ polygonal curves in $\Re^d$
    with total complexity $n$, and let $1 \geq \eps > 0$ and $\delta
    >0$ be two parameters. Then, there is an algorithm
    $\decider(\delta, \eps, \curveA_1,\dots, \curveA_k)$ that, in
    $O\pth{ \SimpComplexityK{\eps}{\curveA_1}{\curveA_k} }$ time, returns
    one of the following outputs:
    \begin{inparaenum}[(i)]
        \item a $(1+\eps)$-approximation to
        $\distMean{\curveA_1}{\curveA_k}$,
        \item $\distMean{\curveA_1}{\curveA_k} < \delta$, or
        \item $\distMean{\curveA_1}{\curveA_k} > \delta$.
    \end{inparaenum}
    
    \lemlab{decider:2}
\end{lemma}

\begin{defn}
    Given a finite set $Z\subseteq \Re$, we say an interval $[\alpha,
    \beta]$ is \emphi{atomic} if it is a maximal interval on the real
    line that does not contain any point of $Z$ in its interior.
\end{defn}

\begin{algorithm}
    For a set of numbers $Z$, let \approxBinarySearch{}$(Z, \eps,\curveA_1,\dots,\curveA_k)$ denote the
    algorithm that performs a binary search over the values of $Z$, to
    compute the atomic interval of $Z$ that contains $\distMean{\curveA_1}{\curveA_k}$. 
    This procedure would use $\deciderFr$ (\lemref{decider:2}) to perform the decisions during
    the search.

    \alglab{a:binary:search}
\end{algorithm}

\begin{lemma} 
    Given a set $\PntSet$ of $n$ points in $\Re^d$, let
    $\Pairwise{\PntSet}$ be the set of all pairwise distances of
    points in $\PntSet$.  Then, one can compute in $O(n \log n)$ time
    a set $Z$ of $O(n)$ numbers, such that for any $y \in
    \Pairwise{\PntSet}$, there exist numbers $x,x' \in Z$ such that $x
    \leq y \leq x' \leq 2x$.  Let $\approxDistances(\PntSet)$ denote
    this algorithm.
    
    \lemlab{all:distances}
\end{lemma}

The following subroutine, from \cite{dhw-afdrc-10}, will allow us to efficiently check intervals with bounded spread for $\distMean{\curveA_1}{\curveA_k}$.

\begin{lemma}
    Given $k$ curves $\curveA_1,\dots,\curveA_k$ in $\Re^d$ of total
    complexity $n$, a parameter $1 \geq \eps>0$, and an interval
    $[\alpha,\beta]$, one can compute a $(1+\eps)$-approximation to
    $\distMean{\curveA_1}{\curveA_k}$ if $\distMean{\curveA_1}{\curveA_k} \in
    [\alpha,\beta]$, or report that $\distMean{\curveA_1}{\curveA_k} \notin
    [\alpha,\beta]$.  The algorithm, denoted by \intervalFr{}
    $([\alpha, \beta], \eps,$ $\curveA_1,\dots, \curveA_k)$, takes $\ds O\pth{
    \SimpComplexityK{\eps}{\curveA_1}{\curveA_k} \log \frac{\log
          (\beta/\alpha)}{\eps} }$ time.

    \lemlab{Frechet:naive}
\end{lemma}

We will also need a new subroutine, called \solver, but first we prove the following easy 
lemma about \MST's.

\begin{lemma}
    Let $\Graph$ be a graph with non-negative weights on its
    edges. For any two vertices $u,v \in \Vertices{\Graph}$,
    for the unique path $\curveC$ between $u$ and $v$ in the \MST, we have
    that $\elevX{\curveC} \leq \elevX{\curveB}$, where $\curveB$ is
    any $uv$ path in $\Graph$, and $\elevX{\curveC}$ is the maximum
    weight edge along the path $\curveC$.

    \lemlab{easy}
\end{lemma}

\begin{proof:in:appendix}{\lemref{easy}}
    For the sake of simplicity of exposition assume that all the
    weights on the edges of $\Graph$ are distinct.%

    Consider a $u v$ path $\curveB$ in $\Graph$. If $\curveB$ is
    contained in the \MST then we are done. Otherwise, let $\edge$ be
    any edge of $\curveB$ that is not contained in the
    \MST. Introducing the edge $\edge$ into the \MST creates a cycle,
    where all the other edges on the cycle are lighter than $\edge$
    (otherwise, $\edge$ must be in the \MST). Therefore, we can replace
    $\edge$ in $\curveB$ by the portion of this cycle connecting its
    endpoints. This new path $\curveB'$ has one less edge outside the
    \MST, and it holds that $\elevX{\curveB'} \leq
    \elevX{\curveB}$. Continuing in this fashion, we end up with a
    path $\curveC'$ in the \MST between $u$ and $v$ such that
    $\elevX{\curveC'} \leq \elevX{\curveB}$. Since the path
    in the \MST between $u$ and $v$ is unique, the claim now follows.
\end{proof:in:appendix}

\begin{lemma}
	Let $\curveA_1, \dots, \curveA_k$ be $k$ polygonal curves in $\Re^d$
   with total complexity $n$, $1 \geq \eps > 0$ be a given parameter, 
	$\delta^*=\distMean{\curveA_1}{\curveA_k}$, and 
	$\SimpComplexityC =\SimpComplexityK{\eps}{\curveA_1}{\curveA_k}$.
	Let $[\alpha,\beta]$ be an atomic interval that contains $\delta^*$,
   and such that for any $\mu, \mu' \in [\alpha,\beta]$, $\simpX{\curveA_i,  
   \mu}=\simpX{\curveA_i, \mu'}$ for $i=1,\dots, k$. 
	Then one can compute in $O(\SimpComplexityC\log \SimpComplexityC)$
	time, a value $\delta$ such that $\delta^*\in [\delta-2\alpha, \delta+2\alpha]$.
   Let this algorithm be denoted by \solver{}$([\alpha,\beta],\curveA_1,\dots,\curveA_k)$

	\lemlab{decider2}
\end{lemma}

\begin{proof}
Let $\mu=\beta$.  We run the algorithm of \lemref{easy:2} on  
$\curveA_1' = \simpX{\curveA_1,\sRadius}$, $\dots$,  $\curveA_k' = \simpX{\curveA_k,\sRadius}$,
except with the following modifications.
First, instead of using the bottleneck shortest path algorithm of \lemref{bottleneck}, 
we will use Prim's algorithm, starting from the vertex that corresponds to the starting 
points of the curves, where we stop when we reach the vertex that 
corresponds to the ending points of the curves.  Also, instead of explicitly 
computing the cell graph, we only compute the relevant parts of the cell 
graph on the fly as they are needed for Prim's algorithm. 
Note that if $\delta$ is the elevation of the shortest path in the \MST from $s$ to $t$,
then Prim's is guaranteed to stay within $\Nleq{\delta}(\curveA_1',\dots,\curveA_k')$
until reaching $t$.  Moreover, by \lemref{easy} this path will be the lowest $st$ path.
 
This modified version of the algorithm computes a curve $\curveB$ 
that minimizes $\max_{i} \distFrW{\curveA_i'}{\curveB}$, in $O(\SimpComplexityC\log \SimpComplexityC)$ time, 
since we are running Prim's on an effective graph of size $\SimpComplexityC$ (and where $E(G)=O(V(G))$).  
Observe that since the $\mu$ simplification is constant on the interval $[\alpha,\beta]$, $\delta$ is the same value that would be returned had we set $\mu=\delta^*$.  Also, again since the $\mu$ simplification is constant on this interval, by \obsref{strong:weak} and considering $\mu=\alpha$, we know that $\delta^*\in [\delta-2\alpha, \delta+2\alpha]$.
\end{proof}

\subsection{Algorithm}
\seclab{algCpack}
\begin{figure}[t]
    \begin{center}
        \fbox{~~~~~~\begin{minipage}{0.90\linewidth}
               \hspace{-0.7cm}\approxMean{}($\eps$, $\curveA_1$, $\dots$,$\curveA_k$)
               \begin{compactenum}[(A)]
                   \item \itemlab{points}
                   $\PntSet = \VertexSet{\curveA_1} \cup \dots \cup
                   \VertexSet{\curveA_k}$
 
                   \item \itemlab{apxd}
                   $\sEvents \leftarrow \approxDistances(\PntSet)$
                   (\lemref{all:distances}).
                  
                   \item \itemlab{searchEvents}
                   $[\alpha,\beta] \leftarrow \approxBinarySearch(
                   Z, \eps, \curveA_1, \dots, \curveA_k)$
                   (\algref{a:binary:search}).
            
                   \item \itemlab{int1}
                   Call \intervalFr{}$\pth[]{[\alpha, 8\alpha], \eps,
						 \curveA_1,\dots,\curveA_k}$
                   (\lemref{Frechet:naive}).
                   
                   \item \itemlab{int2}
                   Call \intervalFr{}$\pth[]{[\beta/2, \beta], \eps,
						 \curveA_1,\dots,\curveA_k }$. 
    
                   \item \itemlab{solver}
						 $\delta \leftarrow \solver{}([2\alpha,\beta/2],\curveA_1,\dots,\curveA_k)$
						 (\lemref{decider2}).

						 \item \itemlab{int3}
						 Return the value returned by \intervalFr{}$\pth[]{[\delta/2, 3\delta/2], \eps,
						 \curveA_1,\dots,\curveA_k }$.

               \end{compactenum}
           \end{minipage}~~}
    \end{center}
    \vspace{-0.6cm}
    \caption{The basic approximation algorithm.}
    \figlab{algorithm}
\end{figure}

Given $k$ curves, $\curveA_1,\dots,\curveA_k$, \figref{algorithm} shows 
the algorithm to efficiently compute  a $1+\eps$ approximation to 
$\distMean{\curveA_1}{\curveA_k}$.  Note that the algorithm depicted in
\figref{algorithm} performs numerous calls to \deciderFr, with an 
approximation parameter $\eps>0$.  If any of these calls discovers the
approximate distance, then the algorithm immediately stops and returns
the approximation. As such, at any point in the execution of the
algorithm, the assumption is that all previous calls to \deciderFr
returned a direction where the optimal distance must lie.

\subsection{Correctness and Running Time}

\subsubsection{Correctness}
In order to apply the algorithm of \lemref{decider2} we first need to find an atomic interval (or subinterval), $[\alpha, \beta]$, that contains $\delta^*=\distMean{\curveA_1}{\curveA_k}$, such that none of the $\mu$ simplifications of any of the $k$ curves change for any choice of $\mu \in [\alpha, \beta]$.  Note that by the way in which $\mu$ simplified curves are constructed, \algref{simplification:algorithm}, if we consider increasing the value of $\mu$ from $0$ to $\infty$, the only events at which any of the $\mu$ simplifications of any of the curves change, are when $\mu$ is equal to one of the distances between a pair of vertices on one of the curves.  Hence if $Y$ denotes the set of all pairwise distances between vertices in $P$ (step \itemref{points} in the algorithm) then in order to apply \lemref{decider2} we want the atomic interval with respect to $Y$ that contains $\delta^*$.  Since it is costly to compute $Y$ explicitly, we instead compute an $O(n)$ sized set $Z$ (step \itemref{apxd}), such that each value in $Y$ is 2-approximated by some value in $Z$.   Step \itemref{searchEvents} performs a binary search over $Z$, using \decider, in order to find an atomic interval $[\alpha, \beta]$ containing $\delta^*$.  Since each value in $Y$ is 2-approximated by some value in $Z$, we know that the interval $[2\alpha, \beta/2]$ is a subinterval  of an atomic interval of $Y$.  Hence by \lemref{Frechet:naive} we know that steps \itemref{int1} and \itemref{int2} ensure that $[2\alpha, \beta/2]$ is a subinterval of an atomic interval of $Y$ that contains $\delta^*$ (and if not, these steps returned a $(1+\eps)$-approximation for $\delta^*$).    
By \lemref{decider2} we know that in step \itemref{solver}, when we call \solver on the interval $[2\alpha,\beta/2]$ we get a value $\delta$ such that $\delta^* \in [\delta-4\alpha,\delta+4\alpha]$.  However, \itemref{int1} guaranteed that $\delta\geq 8\alpha$, since we checked the interval $[\alpha, 8\alpha]$.  This implies $\alpha\leq \delta/8$ and so $\delta^*\in [\delta-4\alpha, \delta+4\alpha]$ implies that $\delta^* \in [\delta-\delta/2, \delta+\delta/2]$.  Hence we have an interval with bounded spread which contains $\delta^*$ and so by \lemref{Frechet:naive}, \itemref{int3} efficiently computes a $(1+\eps)$-approximation for $\delta^*$.

\subsubsection{Running Time}

Let $n=|P|$ and $\SimpComplexityC= \SimpComplexityK{\eps}{\curveA_1}{\curveA_k}$.  By \lemref{all:distances} that the call to \approxDistances in \itemref{apxd} takes $O(n\log n)$ time.  Since \approxBinarySearch just preforms a binary search over the $O(n)$ values returned by \approxDistances by using \decider, which runs in $O(\SimpComplexityC)$ time by \lemref{decider:2}, we know that \itemref{searchEvents} takes $O(\SimpComplexityC \log n)$ time.  Since $[\alpha,8\alpha]$, $[\beta/2,\beta]$, and $[\delta/2, 3\delta/2]$ are all intervals with bounded spread, we have by \lemref{Frechet:naive}, that steps \itemref{int1}, \itemref{int2}, and \itemref{int3} run in $O(\SimpComplexityC \log (1/\eps))$ time.  Finally, by \lemref{decider2}, the call to \solver in line \itemref{solver} takes $O(\SimpComplexityC \log \SimpComplexityC)$ time.  We thus have the following.

\begin{lemma}
\lemlab{algCpackTheorem}
Let $\curveA_1, \dots, \curveA_k$ be $k$ given polygonal $c$-packed curves in $\Re^d$ of total complexity $n$, let $\eps>0$ be a parameter, and let $\SimpComplexityC= \SimpComplexityK{\eps}{\curveA_1}{\curveA_k}$.  Then one can compute, in $O(\SimpComplexityC \log (n/ \eps)+ n\log n)$ time, reparameterizations of the curves that $1+\eps$ approximate the value of $\distMean{\curveA_1}{\curveA_k}$.  In particular, one can $1+\eps$ approximate the mean curve of $\curveA_1, \dots, \curveA_k$.
\end{lemma}

\subsubsection{Free Space Complexity}

\begin{lemma}
    For $k$ $c$-packed curves $\curveA_1,\dots,\curveA_k$ in $\Re^d$
    of total complexity $n$, and $0 < \eps < 1$, we have that
    $\SimpComplexityC=\SimpComplexityK{\eps}{\curveA_1}{\curveA_k} = O( (c /\eps)^{k-1}n)$.
    
    \lemlab{complexity:l:e:q}
\end{lemma}

\begin{proof}
    Let $\delta \geq 0$ be a fixed parameter, $\sRadius =\eps
    \delta$, and $\curveA_1' = \simpX{\curveA_1,\sRadius}$, $\dots$,  
    $\curveA_k' = \simpX{\curveA_k,\sRadius}$.

The free space in the polyhedral complex is partitioned into connected components.  We must bound the size of the component which contains the start vertex, that is the reachable free space, $\Reachable$.  Observe that one can charge a maximal dimensional cell in the polyhedral complex to an adjacent lower dimensional cell, since maximal cells contribute to $\SimpComplexityC$ only if one of their adjacent proper subcells contributes to $\SimpComplexityC$.
A non-maximal cell corresponds to some vertex $v$ on one of the curves, and either a vertex or an edge from each one of the $k-1$ other curves.  Consider a ball, $\BallC$, of radius $r=2\delta$ centered at $v$.  We now wish to count the number of features from the other curves (i.e. edges or vertices) that intersect this ball.

\parpic[r]{\includegraphics[scale=.5]{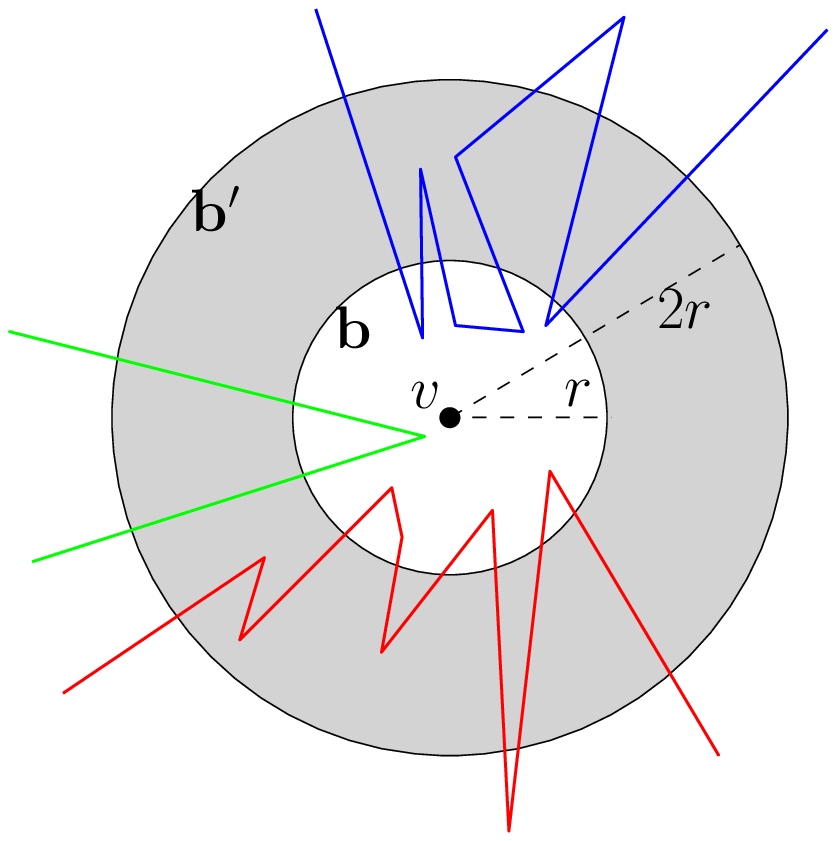}}

To this end, consider one of the other curves, $\curveA_i '$.  Let
$X_i$ be the set of all features of $\curveA_i'$ that intersect
$\BallC$.  Consider a ball, $\BallC '$, of radius $2r$ around
$v$. Since $r\geq \mu$ and the edges of a $\mu$ simplified curve are
of length $\geq \mu$ (with the exception of the last edge), every edge
feature in $X_i$ must contribute at least length $\mu$ to the
intersection of $\BallC'$ and $\curveA_i'$ (note that if the feature
is a vertex, then it is adjacent to an edge which contributes at least
length $\mu$).  By \lemref{simplification:distance}, the total length
of $\curveA_i '$ inside this $\BallC '$ is at most $12cr$.  Therefore,
    \[
    |X_i|= O\pth{\frac{\distCmd{ \curveA_i'\cap \BallC '}}{\mu}} = O\pth{\frac{c r}{\mu}} = O\pth{\frac{c \delta}{\eps \delta}} = O\pth{\frac{c}{\eps}}.%
    \]
Similarly, for each of the other $k-1$ simplified curves, there are also $O(c/\eps)$ features close enough to $v$, that can be involved in a cell that contributes to $\SimpComplexityC$. Such a cell in the polyhedral complex involves choosing the vertex $v$, and one of these $O(c/\eps)$ features from each of the other $k-1$ curves, and hence there are $|X_1|\cdot|X_2|\ldots \cdot|X_{k-1}| =O((c/\eps)^{k-1})$ such cells.  Since there are $n$ vertices in total we thus have that $\SimpComplexityC=O((c/\eps)^{k-1}n)$. 
\end{proof}

\subsection{The Result}

\begin{theorem}
Let $\curveA_1, \dots, \curveA_k$ be $k$ given polygonal $c$-packed curves in $\Re^d$ with total complexity $n$, let $\eps>0$ be a parameter, and let $\SimpComplexityC= \SimpComplexityK{\eps}{\curveA_1}{\curveA_k} =O((c/\eps)^{k-1}n)$.  Then one can compute, in $O(\SimpComplexityC\log \SimpComplexityC)$ time, reparameterizations of the curves that $(1+\eps)$-approximate the value of $\distMean{\curveA_1}{\curveA_k}$.  In particular, one can $1+\eps$ approximate the mean curve of $\curveA_1, \dots, \curveA_k$ in $O(\SimpComplexityC\log \SimpComplexityC) = O((c/\eps)^{k-1}n\log n)$ time.
\end{theorem}

\section{Computing Optimal \Frechet Paths for \DAG Complexes}
\seclab{DAG:complexes}

In this section, we present a simple algorithm for computing exactly
the monotone \Frechet distance between two polygonal curves. This
algorithm has running time $O\pth{n^2 \log n}$, and uses
randomization instead of parametric search.  In fact, the algorithm is
considerably more general and applies to a wider class of inputs.

\paragraph{\DAG complexes.}
Consider a directed acyclic graph (\DAG) with vertices in $\Re^d$,
where a directed edge $\pnt\rightarrow\pntA$ is realized by the
segment $\pnt\pntA$. We refer to such a graph as being a \emphi{\DAG
   complex}. Given two \DAG complexes $\dcA$ and $\dcB$,
start vertices $s_1 \in \Vertices{\dcA},s_2 \in
\Vertices{\dcB}$, and end vertices $t_1 \in
\Vertices{\dcA}, t_2 \in \Vertices{\dcB}$, 
the problem is finding two directed polygonal paths
$\curveA_1, \curveA_2$ in $\dcA$ and $\dcB$, respectively,
such that:
\begin{compactenum}[(A)]
    \item The path $\curveA_i$ uses only edges that appear in
    $\complex_i$, and it traverses them in the direction compliant
    with the orientation of the edges in $\complex_i$, for $i=1,2$.

    \item The curve $\curveA_i$ connects $s_i$ to $t_i$ in
    $\complex_i$, for $i=1,2$.

    \item The monotone \Frechet distance between $\curveA_1$ and
    $\curveA_2$ is minimized among all such curves.
\end{compactenum}

\medskip

Note that this problem includes the problem of computing the
monotone \Frechet distance between two polygonal curves (i.e., 
orient the edges of the curves in the natural way and consider them to
be \DAG complexes).

\subsection{The Decision Procedure}

The algorithm is a direct extension of the work of \cite{ag-cfdbt-95}.
Their algorithm relied on the fact that there was a clear topological
ordering on the cells of the free space, and hence reachability
information could be propagated.  In this case, there is also a topological
ordering (since it is a \DAG).  Hence, in the product space of two \DAG
complexes there is an ordering of the cells according to the underlying
ordering of the two \DAG{}s, and this ordering is acyclic. 

So, let $\dcA$ and $\dcB$ be the two given \DAG
complexes,  $\delta$ a specified radius, and $s_1,s_2, t_1, t_2$ 
the given vertices. The problem is to decide if there are paths 
between the start and end vertices in the corresponding complexes 
of \Frechet distance at most $\delta$.

\paragraph{Algorithm.} 

Compute the topological orderings of the cells (i.e., vertices and
edges) of $\dcA$ and $\dcB$.  In the resulting ordering
$\prec_i$, it holds that $\cell \prec_i \cell'$ if $\cell$ appears
before $\cell'$ in this ordering, for $i=1,2$, where $\cell, \cell' \in \complex_i$.

We compute the product complex $\complex = \dcA \times
\dcB$, and compute the topological ordering of the cells of
$\complex$. Formally for $\cell = \pth{\cell_1, \cell_2}, \cell' =
\pth{\cell_1', \cell_2'} \in \complex$ we have that $\cell \preceq
\cell'$ if and only if $\cell_1 \preceq_1 \cell_1'$ and $\cell_2
\preceq_2 \cell_2'$. Clearly, the ordering $\prec$ over the cells of
$\complex$ is acyclic, and can be computed in
linear time in the size of the complex.

Now, just as in \cite{ag-cfdbt-95}, we start at the start vertex in
the product space $(s_1, s_2)$, visit cells according to their
topological order, and compute the free space and propagate
reachability information on the fly when we reach a new cell.

Since we are working in the product space instead of in the parametric
space, the two dimensional cells are parallelograms instead of squares.

The reachability information is being propagated in a manner similar
to \cite{ag-cfdbt-95}, except we propagate between adjacent cells, 
instead of neighboring two dimensional cells.  Note, that no pair of two
dimensional cells are directly adjacent, as there must be a one
dimensional cell separating them. As such, for each edge (i.e., one
dimensional cell) of $\complex$ we maintain the set of reachable
points.  Unlike in \cite{ag-cfdbt-95}, the reachability information
along a bounding edge in the product space might not be a single
interval, since potentially multiple cells propagate to that bounding
edge.  However, by \lemref{convex}, we only need to compute
the first point (according to the ordering along this edge) that is
reachable on this edge (notice, that an edge is always a product of a
vertex of one curve and a directed edge of the other curve, and as
such it has a natural ordering). 

In particular, when the algorithm visits a cell $\cell$ in this
ordering, it fetches all the cells that
are adjacent to it and appear before it in the ordering. For each adjacent
cell, the reachability information computed is of constant size, and hence we can
compute the reachability information for the new cell in constant
time. Indeed, the handling depends on the dimension of $\cell$:
\begin{compactenum}[(A)]
    \item $\dim(\cell)=0$: ($\cell$ is a vertex), the algorithm
    computes if it is reachable from any of its direct ancestors, and if
    so we mark it as reached.

    \item $\dim(\cell)=1$: ($\cell$ is an edge), the algorithm
    computes the first point on the edge reachable from its direct
    ancestors. 

    \item $\dim(\cell)=2$: ($\cell$ is a parallelogram), the algorithm
    uses the reachability information on the two incoming edges and
    the incoming vertex to compute the reachability inside the
    parallelogram. (Clipping the region to $\FDleq{\delta}$ inside
    this cell.)
\end{compactenum}

As the algorithm visits the cells in a topological order, the work in
maintaining the reachability information, can be charged to a cell's
predecessors.  As such, overall, the running time of the algorithm is
linear in the complex size.

The size of a \DAG complex is the number of edges in it (since we
assumed implicitly that the input \DAG complexes are connected).  Let
$n$ be the number of edges in the larger of the two \DAG complexes
under consideration.  There are potentially $O(n^2)$ cells in
the product space $\complex$.   As such, the running time of the
decision procedure is $O(n^2)$.

\begin{lemma}
    Let $\dcA$ and $\dcB$ be two \DAG complexes, $n$ be the number of
    edges in the larger of the two, $s_1,t_1 \in \dcA, s_2, t_2 \in \dcB$
    be start and end vertices, and $\delta\geq0$ be a parameter.
    Then, one can decide, in $O\pth{ n^2}$ time, if there exists two
    paths $\curveA_1$ and $\curveA_2$ in $\dcA$ and $\dcB$,
    respectively, such that
    \begin{inparaenum}[(i)]
        \item $\curveA_i$ connects $s_i$ with $t_i$, for $i=1,2$, and 
        \item $\distFrM{\curveA_1}{\curveA_2} \leq \delta$, where 
              $\distFrM{\cdot}{\cdot}$ is the monotone \Frechet distance.
    \end{inparaenum}
    Furthermore, if such paths exist, the algorithm returns them
    together with their respective reparameterizations realizing this
    distance.

    \lemlab{decider}
\end{lemma}

\subsection{Using the Decision Procedure}

In the following, let $\dcA$ and $\dcB$ be the two \DAG complexes
under consideration.  We outline a randomized algorithm to compute the
value of the \Frechet distance between the two curves in $\dcA$ and
$\dcB$, that start and end at their respective start and end vertices,
that minimize the \Frechet distance. 

The algorithm needs to search over the critical values when the
decision procedure changes its behavior. These critical values are the
same as in Alt and Godau \cite{ag-cfdbt-95} (vertex-vertex,
vertex-edge and monotonicity events).  Indeed, for any pair of paths
in the \DAG complexes, the critical values for these two paths are the same as
in \cite{ag-cfdbt-95}. As such, since \DAG complexes are the union of paths,
the critical values are the same.

In the following, let $\delta^*$ denote the actual minimum value of
the \Frechet distance.  Given a parameter $\delta$, let
\decider{}($\delta$) be the decision procedure described above.  Let
\extractI{}$(a,b)$ be a procedure that returns all critical values
determined by $\dcA$ and $\dcB$ whose radius is in the interval $[a,
b]$.  Suppose, for the time being, that the following subroutines have
the following running times:
\medskip
\begin{compactenum}[(A)]
    \item \decider{}($\delta$) runs in $O(n^2)$ time
    (\lemref{decider}).

    \item \extractI{}$(a,b)$ runs in $O(n^2\log n + k\log n)$ time,
    where $k$ is the number of critical values with radius in the
    interval $[a,b]$.

    \item One can uniformly sample a critical value from the set of all
    critical values in $O(1)$ time per sample.
\end{compactenum}

\medskip

\begin{figure}
    \centerline{\fbox{
          \begin{minipage}{0.5\linewidth}%
              \begin{tabbing}
                  \ \ \ \ \ \ \ \= \ \ \ \ \ \ \ \ \= \ \  \kill
                  \compFr{}($\dcA,\dcB, \sA, \sB, \tA, \tB$): \\
                  \> $R$: random sample of $\mu = 4 n^2$ critical values \\
                  \> Sort $R$ \\
                  \> Perform a binary search over $R$ using \decider\\
                  \> $\Interval = [a,b]  \leftarrow$ Atomic interval of $R$
                  containing $\delta^*$\\
                  \> $S \leftarrow$ \extractI{}$(a,b)$ \+\\
                  \>  \texttt{// $S$: all critical values in
                     $[a,b]$} \-\\
                  \> Sort $S$\\
                  \> $x \leftarrow$ Smallest value in $S$ for which \decider
                  accepts \+\\
                  \> \texttt{// Computed using a binary search} \-\\
                  \> Return $x$
              \end{tabbing}
          \end{minipage}
       }
    }
    \caption{The algorithm for computing the \Frechet distance between
       two \DAG complexes.}
    \figlab{f:r:algorithm}
\end{figure}

The new algorithm is depicted in \figref{f:r:algorithm}.

\subsubsection{Computing the Critical Values in an Interval}

To complete the description of the algorithm, we need to describe how
to implement \extractI{}$(a,b)$.  For the interval $\Interval =
[a,b]$, we need to compute all the critical values with radius in
$\Interval$.  We can explicitly compute all the radii of vertex-vertex
and vertex-edge events in this interval and sort them in $O(n^2\log
n)$ time, where $n$ is the number of edges (since there are
$O\pth{n^2}$ such events in total and each radius can be computed in
$O(1)$ time). Indeed, for a vertex-vertex event, its radius is the
distance between the two vertices that define it. Similarly, the radius
of a vertex-edge event is the distance between a vertex and an
edge. Both types of radii can be computed in constant time, given
the two elements that define them.

In order to compute the radii of monotonicity events in $\Interval$,
we apply a variant of the standard line sweeping algorithm (i.e \KDS).
Specifically, for two \DAG complexes $\dcA$ and $\dcB$, consider
finding all monotonicity events between an edge $\edge$ of $\dcA$, and
pairs of vertices from $V=\Vertices{\dcB}$. To this end, place a
sphere of radius $\delta$ at each point of $V$ with radius $\delta=a$.
We now increase the radius $\delta$ till it reaches $b$.  The
algorithm maintains an ordered list $L$ of the intersections of the
spheres with the edge $\edge$.  The events in this growing process
are:
\begin{compactenum}[(A)]
    \item The first time a sphere intersects $\edge$ (this will create
    two intersections, if the intersection happens internally on
    $\edge$, since after this point the sphere will intersect $\edge$
    in two places).

    \item When the intersection point of a sphere with $\edge$ grows past
    an endpoint of $\edge$.

    \item When two different spheres intersect at the same point on
    $\edge$.  At this point, the algorithm exchanges the order of 
    these two intersections along $\edge$.
    The value of $\delta$ when such an event happens is the
    radius of a monotonicity event.

\end{compactenum}
\smallskip
At any point in time, the algorithm maintains a heap of future
events. Whenever a new intersection point is introduced, or two
intersections change their order along $\edge$, the algorithm computes
the next time of an event involving these intersections with the
intersections next to them along $\edge$.

It is clear that between such events the ordering of the intersections
of the spheres with $\edge$ does not change.  Similarly, for a
monotonicity event to happen on $\edge$, there must be a point in time
in which the corresponding spheres are neighbors along $\edge$.
Hence, this algorithm will correctly find all the monotonicity events.

It takes $O\pth{(n+k)\log n}$ time to compute all the relevant
monotonicity events involving $\edge$ and $V$, where $k$ is the number
of such events.  We must do this for all edges of $\dcA$ and hence it
takes $O((n^2+k')\log n)$ time to compute all the monotonicity events
between edges of $\dcA$ and vertices of $\dcB$, where $k'=\sum_i k_i$
and $k_i$ is the number of monotonicity events in the interval $[a,b]$
involving the $i$\th edge of $\dcA$.  Therefore it takes
$O((n^2+k'')\log n)$ time to compute all the relevant monotonicity
events between $\dcA$ and $\dcB$, where $k''$ is the number of such
events (i.e. both those involving edges of $\dcA$ and those involving
edges of $\dcB$).

\subsubsection{Sampling Critical Values}

We can uniformly sample critical values in $O(1)$ time, as follows.  A
vertex-edge event is determined by sampling a vertex and an edge, a
vertex-vertex event is determined by sampling a pair of vertices, and
a monotonicity event is determined by sampling a pair of vertices and
an edge.  Since we can easily uniformly sample vertices and edges in
$O(1)$ time, we can therefore do so for critical events.  In general,
the decision of which type of critical event to sample would have to
be weighted by the respective number of such events.

\subsection{Analysis}
Let $R$ be the random sample of critical values, of size $O(n^2)$.
The interval $[a,b]$ computed by \compFr contains $\delta^*$.  The
call to \extractI{}$(a,b)$ takes $O(n^2\log n + k\log n)$ time, where
$k$ is the number of monotonicity events.  The following lemma shows
that $k=O(n^2)$.

\begin{lemma}
    Let $\Interval = [a,b]$ be the interval computed by
    \compFr{}, and let $c$ be some positive constant.  Then,
    \[
    \Pr \pbrc{\text{number of critical events in }[a, b] > 2c n\ln
       n} \leq \frac{1}{n^c}.
    \]    
\end{lemma}

\begin{proof}
    There are $2\binom{n}{2} n \leq n^3$ possible monotonicity events, $2n^2$
    possible vertex-edge events, and $n^2$ possible vertex-vertex
    events. As such, the total number of critical events is bounded by
    $Z = n^3 + 2n^2 + n^3 \leq 2 n^3$.

    Consider the position of $\delta^*$ on the real line.
    Let $C$ be the set of the radii of all these critical events,
    and let $U^-$ (resp. $U^+$) be the set of $M = cn\ln n$ values of
    $C$ that are closest to $\delta^*$ that are smaller (resp. larger)
    than it, and let $U = U^- \cup U^+$.
 
    If the number of values in $C$ smaller than $\delta^*$ is at most
    $M$, then there could be at most $M$ critical values smaller than
    $\delta^*$ in $[a,b]$. The same holds if the $C$ contains less
    than $M$ values larger than $\delta^*$. As such, in the following,
    assume that both quantities are larger than $M$.

    The probability that the random sample $R$ of size $\mu=4 n^2$
    picked by the algorithm, does not contain a point of $U^-$, is at
    most
    \[
    \pth{1-\frac{\cardin{U^-}}{\cardin{C}} }^\mu 
    \leq 
    \pth{1-\frac{c\ln n}{2 n^2}}^{4 n^2} \leq
    \exp\pth{-2 c\ln n} \leq \frac{1}{2 n^c}.
    \]
    This also bounds the probability that $R$ does not contain a value
    of $U^+$. As such, with high probability, $[a,b]$ contains only
    events in the set $U$. Namely, $[a,b]$ contains the radii of at
    most $\cardin{U^-} + \cardin{U^+} \leq 2M$ monotonicity events,
    with probability $\geq 1- 1/n^c$.
\end{proof}

Combining all our results, we thus have the following theorem.
\begin{theorem}
    For two \DAG complexes, $\dcA$ and $\dcB$, of total
    complexity $n$, with start and end
    vertices $\sA,\tA \in \dcA, \sB, \tB \in \dcB$, 
    the algorithm \compFr{}$(\dcA, \dcB, \sA, \tA,
    \sB, \tB)$ returns two curves $\curveA_1$ and $\curveA_2$, such
    that $\curveA_1$ (resp. $\curveA_2$) connects $\sA$ (reps. $\sB$)
    to $\tA$ (resp. $\tB$) in $\dcA$ (resp. $\dcB$).  Furthermore,
    the monotone \Frechet distance between $\curveA_1$ and
    $\curveA_2$, is the minimum among all such curves.  The running
    time of the algorithm is $O\pth{ n^2\log n }$, with
    probability $\geq 1-1/n^c$.

    \thmlab{f:r:DAG}
\end{theorem}

\begin{remark}
    The above result implies that given two polygonal curves in
    $\Re^d$ one can compute the \Frechet distance between them, in
    $O(n^2 \log n)$ time (this running time bound holds with high
    probability), by a simple algorithm that does not use parametric
    search.
\end{remark}

\section{Conclusions}

In this paper, we showed that the algorithm for computing the (weak)
\Frechet distance between two curves can be extended to more general
settings. This results in a slew of problems that can be solved using
the new framework.

\paragraph{Monotonicity.} 
Our main algorithm from \secref{alg} is an extension of the algorithm of Alt and Godau
\cite{ag-cfdbt-95} for the weak \Frechet distance. It is natural to
ask if the new framework can handle monotonicity. 
In \secref{DAG:complexes}, we offered a very restricted extension of  
our framework to this case, in the process presenting a new simpler
algorithm for computing the monotone \Frechet distance between
polygonal curves.

For more general settings, if the underlying complex is not one
dimensional then it is not clear what monotonicity means. Even if we
restrict ourselves to the case of $k$ input curves, for $k>2$, it is
not immediately clear how to handle monotonicity efficiently, and we
leave this as an open problem for further research. Interestingly,
there are cases where monotonicity actually makes the problem easier.

\paragraph{Running Time.}
The running time of the general algorithm is $O\pth{ n^k }$
when handling $k$ input complexes and is probably practical only for
very small values of $k$.  In \secref{pack} we showed that one can get a 
$(1+\eps)$-approximation for the mean curve problem for $k$ $c$-packed 
curves in $\widetilde{O}(n\log n)$ time.  It should be possible to extend 
this same procedure to approximate, in a similar running time, some of the other 
problems that are solved by the general framework, under similar assumptions on the input.

\subsection*{Acknowledgments.}

The authors thank Anne Driemel, Jeff Erickson, Steve LaValle, Jessica
Sherette and Carola Wenk for useful discussions on the problems
studied in this paper.

\bibliographystyle{alpha}
\bibliography{shortcuts,geometry,new}

  \immediate\closeout\myoutfile. 
  \section{Proofs}
  \input{fragment/myfile.tmp}

\end{document}